\documentclass[aps,prb,reprint]{revtex4-1}

\usepackage{graphicx}
\usepackage{amsmath}
\usepackage{amsfonts}
\usepackage{amssymb}
\usepackage{bbm}
\usepackage[]{units}
\usepackage{hyperref}
\usepackage{braket}

\newcommand{\nnb}{\nonumber \\}
\newcommand{\im}{\mathrm{i}}
\newcommand{\bv}{\left( \begin{array}{c}}
\newcommand{\ev}{\end{array} \right)}
\newcommand{\E}{\mathrm{e}}
\newcommand{\ko}{\text{, }}

\newcommand{\st}[1]{_{\text{#1}}}
\newcommand{\bo}[1]{\boldsymbol{#1}}
\newcommand{\I}{\mathrm{i}}

\newcommand{\braketv}[2]{\left\langle #1\right|\left.\! #2 \right\rangle }
\begin{document}
\title{Long distance coupling of resonant exchange qubits}
\author{Maximilian Russ and Guido Burkard}
\affiliation{Department of Physics, University of Konstanz, D-78457 Konstanz, Germany}
\begin{abstract}
We investigate the effectiveness of a microwave cavity as a mediator of interactions between two resonant exchange (RX) qubits in semiconductor quantum dots (QDs) over long distances, limited only by the extension of the cavity. Our interaction model includes the orthonormalized Wannier orbitals constructed from Fock-Darwin states under the assumption of a harmonic QD confinement potential. We calculate the qubit-cavity coupling strength in a Jaynes Cummings Hamiltonian, and find that dipole transitions between two states with an asymmetric charge configuration constitute the relevant RX qubit-cavity coupling mechanism. The effective coupling between two RX qubits in a shared cavity yields a universal two-qubit i\textsc{swap}-gate with gate times on the order of nanoseconds over distances on the order of up to a millimeter.
\end{abstract}
\maketitle
%%%%%%%%%%%%%%%%%%%%%%%%%%%%%%%%%%%%%%%%%

\section{Introduction}

Quantum computation with single-electron spins confined in semiconductor quantum dots (QDs) \cite{PhysRevA.57.120} has been investigated within a wide range of implementations yielding long decoherence times on the order of $ \sim\unit{\mu s} $.\cite{Petta2005,Greilich2006,Koppens2008,Bluhm2011} Gallium Arsenide (GaAs)\cite{Hanson2007,Awschalom2013} and silicon (Si)\cite{Zwanenburg2013} are the most common choices of host materials for the QDs. By comparison, charge qubits decohere much faster (within $ \sim \unit{ns} $ \cite{Petersson2010,Shi2013,Kim2015}) due to the strong coupling between the charge and the electromagnetic fields in the environment.\cite{Yurkevich2010} Therefore, the aim of many implementations is a high protection against electrical noise to achieve long qubit decoherence times in order to be able to perform as many qubit operations (gates) as possible while the qubit is coherent. Experiments show high-fidelity state manipulation and long coherence times for scalable implementations using single or multiple QDs.\cite{Gaudreau2006,Takakura2010,Gaudreau2012,Medford2013N,Medford2013,Veldhorst2015,Zajac2015,Otsuka2015} One promising candidate is the resonant exchange (RX) spin qubit, a modification of the exchange-only qubit,\cite{nature2000} which allows for all electric control of the qubit for the price of a triple quantum dot (TQD) scheme. This implementation possesses a high robustness against electrical noise at the sweet spot,\cite{Taylor2013,Fei2015,Russ2015} but is still susceptible to electromagnetic fields at the resonance frequency allowing for additional qubit control through radio-frequency or microwave signals.\cite{Taylor2013,Medford2013}
It was shown that two-qubit gates between RX qubits can be implemented with the exchange coupling,
using a single exchange pulse.\cite{Doherty2013} 

\begin{figure}
		\begin{center}
				\includegraphics[width=1.00\columnwidth]{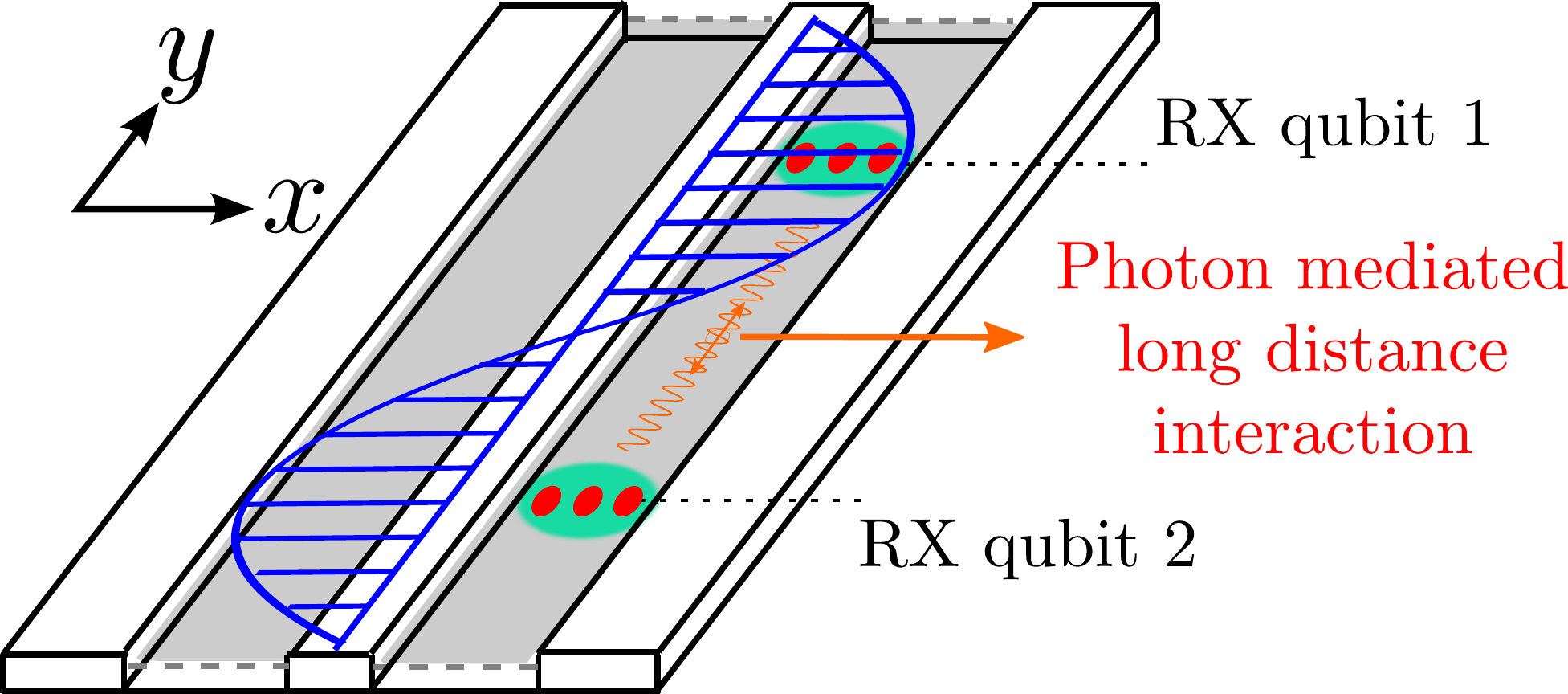}
				\caption{Schematic illustration of the proposed setup for long distance interaction. The setup consists of two linearly arranged TQDs (RX qubits) inside a superconducting strip-line cavity. For an optimal setup the two qubits should be located at the field maxima (anti-nodes) in order to achieve a strong qubit-cavity coupling. The green environment surrounding the RX qubits illustrates the coupling of the qubit to the electromagnetic field of the cavity (blue). As a result, the RX qubit can be coupled by using photons as mediators. The gray layer illustrates the two-dimensional electron gas (2DEG) in which the quantum dots are embedded.}
				\label{fig:model_ULD}
		\end{center}
\end{figure}

The coherent transport of quantum information, e.g., long distance en\-tangle\-ment \cite{Aspelmeyer2003}, combined with state preparation and read-out is investigated within a wide range of implementations \cite{Takeuchi2014} in quantum optics. Adapting techniques from this field,  a long distance coupling between two solid-state QD spin-qubits can be envisioned, which could complement the existing short-range interactions, such as the exchange coupling between nearby QD spin-qubits. In order to achieve long distance coupling between spin-qubits, a long-range interaction is needed, e.g., the coupling of the qubits to an electromagnetic field with specific photon modes \cite{Burkard2006} or by the coupling of the qubit to a ferromagnet.\cite{Trifunovic2013} In this paper, we focus on long-distance coupling between RX qubits based on the electromagnetic fields in a microwave cavity (Fig.~\ref{fig:model_ULD}).

Coupling spin-qubits via electromagnetic cavities has been proposed for spin-qubits in nitrogen vacancy (NV) centers in diamond. NV centers can be coupled with photons in the optical spectrum,\cite{Dobrovitski2013,Burkard2014,Gao2015} while QD spin-qubits usually react to microwave or radio-frequency signals. Since microwave cavities with a high finesse exist,\cite{Xiang2014} this long-range coupling is feasible between QD spin-qubits in solid-state materials, e.g., coupling spin qubits to photons in a cavity via electric dipole or gate potentials.\cite{Burkard2006,Hu2012,Jin2012} Experiments have shown evidence of a strong coupling between qubits in single or double QDs to a microwave cavity through the charge\cite{Petersson2012,Toida2013,Wallraff2013,Liu2014} or the spin.\cite{Frey2012,Basset2013}
Here, we present an implementation of such a long distance interaction between two RX qubits enabled in a TQD scheme at time scales on the order of nanoseconds. 

This paper is organized as follows. In Section \ref{sec:model_ULD} we introduce our model for the long distance interaction. Subsequently, in Section \ref{sec:qubit_cavity} we calculate the associated transition dipole matrix elements used to determine the corresponding qubit-cavity coupling parameter and identify the underlying coupling mechanism. Finally in Section \ref{sec:ULDinter}, we combine these results to present a step-by-step prescription for a universal two-qubit i\textsc{swap}-gate between two RX qubits in a shared cavity. We conclude in Section \ref{sec:conclusion_ULD} with a summary and an outlook.
%%%%%%%%%%%%%%%%%%%%%%%%%%%%%%%%%%%%%%%%%%%%%%%%%%%%%%%%%%%%%%%

\section{Model}
\label{sec:model_ULD}
We consider two linearly arranged triple QDs (TQDs), where each QD has a single available orbital, occupied by three electrons. Both TQDs are assumed to lie in a superconducting microwave cavity with a single available photon mode (see Fig.~\ref{fig:model_ULD}). We describe this system with the Hamiltonian
\begin{align}
	H = \sum_{i=1}^{2}\left( H_i + H\st{int,\textit{i}}\right) + H\st{cav}\ko
	\label{eq:Htotal}
\end{align}
where $ H_i $ describes the dynamics of the electrons in the $ i $-th isolated TQD, $ H\st{int,\textit{i}} $ is the interaction between the electrons in the $ i $-th TQD and the photons in the cavity and $ H\st{cav} $ describes the photons in the cavity. 

\begin{figure}
		\begin{center}
				\includegraphics[width=1\columnwidth]{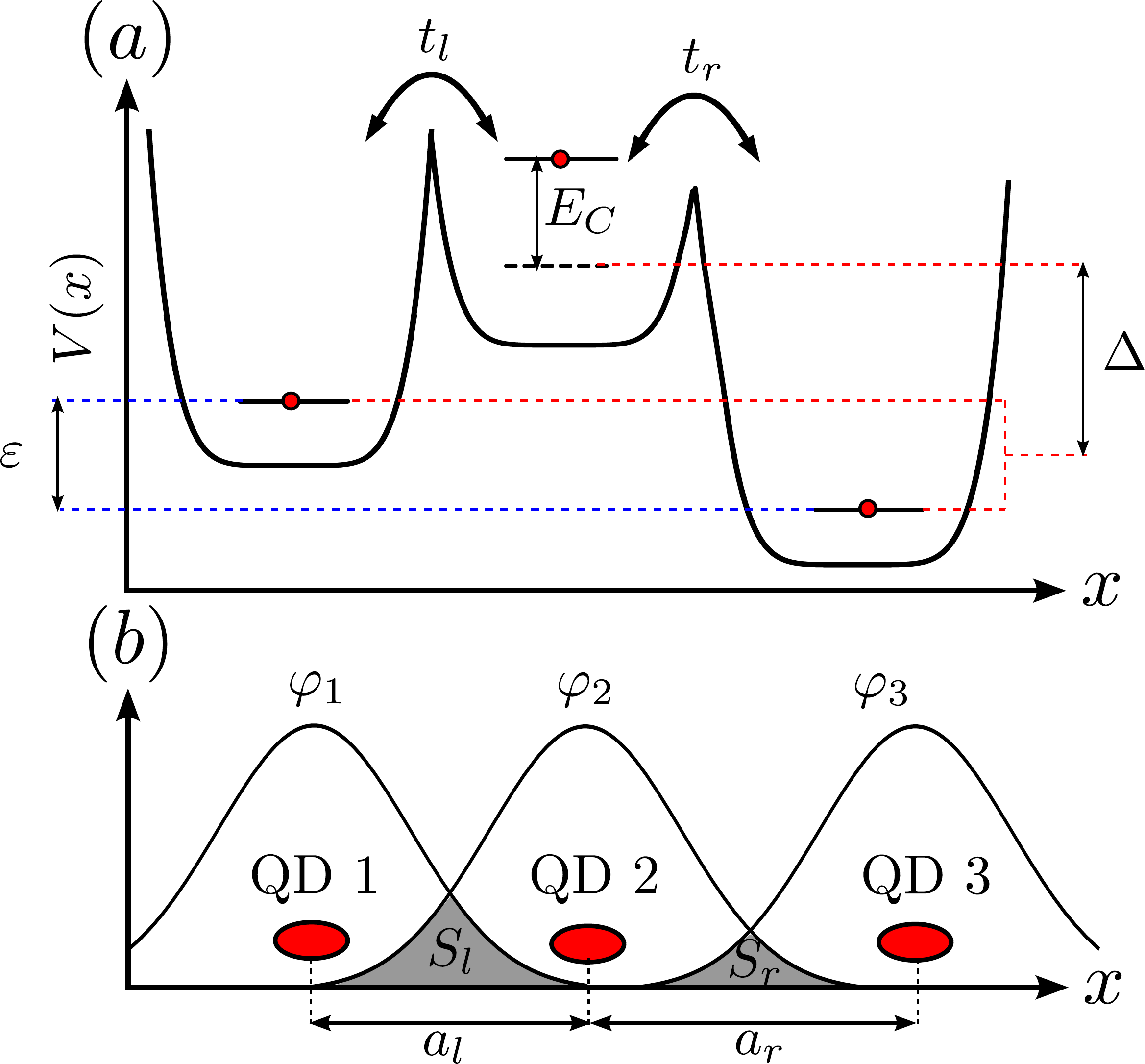}
				\caption{(a) Schematic illustration of the triple quantum dot (TQD) confinement potential $V(x)$. The hopping matrix element between QD~1 (QD~3) and QD~2 is denoted by $t_{l}$ ($t_{r}$). We also show the response of the system to two electrically controlled bias parameters, the difference $\varepsilon$ between the energy levels of the outer two QDs and, the effective difference $\Delta$ of the energy levels in QD~2 and the mean of the outer QDs (including Coulomb repulsion). Here, we assume that the TQD is filled with three electrons. We also include the Coulomb repulsion in the center QD, $E_{C}=U-2U_{C}$. (b) Schematic illustration of the orbital wave functions of the electrons in the TQD. The inter-dot distances $a_{l}$ ($a_{r}$) between QD~1 (QD~3) and QD~2 need to be sufficiently small to allow for a sizeable overlap $S_{l}$ ($S_{r}$) between the orbital wave functions.}
				\label{fig:TQD}
		\end{center}
\end{figure}

For the qubit we use the RX Hamiltonian\cite{Taylor2013,Russ2015} $H_{i}$ derived from the three-site extended Hubbard Hamiltonian which describes a linearly arranged TQD (see Fig.~\ref{fig:TQD}~(a)). We work in the RX regime in which only the charge states (1,1,1), (2,0,1) and (1,0,2) are accessible. The spin qubit lies in the subspace with spin quantum numbers $S=S_{z}=\frac{1}{2}$, spanned by the states 
\begin{align}
	\ket{0}&=\ket{S}_{13}\ket{\uparrow}_2=\frac{1}{\sqrt{2}}\left(c_{1,\uparrow}^\dagger c_{2,\uparrow}^\dagger c_{3,\downarrow}^\dagger-c_{1,\downarrow}^\dagger c_{2,\uparrow}^\dagger c_{3,\uparrow}^\dagger\right)\ket{\text{vac}},   \nnb
	\ket{1}&= \sqrt{\frac{2}{3}}\ket{T_+}_{13}\ket{\downarrow}_2 - \sqrt{\frac{1}{3}}\ket{T_0}_{13}\ket{\uparrow},\nnb
	&=\frac{1}{\sqrt{6}}\!\left(2c_{1,\uparrow}^\dagger c_{2,\downarrow}^\dagger c_{3,\uparrow}^\dagger\!-\!c_{1,\uparrow}^\dagger c_{2,\uparrow}^\dagger c_{3,\downarrow}^\dagger\!-\!c_{1,\downarrow}^\dagger c_{2,\uparrow}^\dagger c_{3,\uparrow}^\dagger\right)\!\ket{\text{vac}},  \nnb	
	\ket{2}&\equiv\ket{S_{1,1/2}} = c_{1,\uparrow}^\dagger c_{1,\downarrow}^\dagger c_{3,\uparrow}^\dagger\ket{\text{vac}},   \nnb 
	\ket{3}&\equiv\ket{S_{3,1/2}} = c_{1,\uparrow}^\dagger c_{3,\uparrow}^\dagger c_{3,\downarrow}^\dagger\ket{\text{vac}}\ko
	\label{eq:logicalspace2}
\end{align}
where $ \ket{\text{vac}} $ denotes the vacuum state and $ c_{i,\sigma}^\dagger $ ($ c_{i,\sigma} $) creates (annihilates) an electron in QD $ i $ with spin $ \sigma $. We have further used the notations $\ket{S}=(\ket{\uparrow\downarrow}-\ket{\downarrow\uparrow})/2$, $\ket{T_{+}}=\ket{\uparrow\uparrow}$, and $\ket{T_{0}}=(\ket{\uparrow\downarrow}+\ket{\downarrow\uparrow})/2$ for the singlet and two of the triplet states of two electrons. Here, $ (m,n,l) $ denote the number of electrons in the first ($ n $), second ($ m $) and third ($ l $) QD. In this basis, we obtain for the Hubbard Hamiltonian\cite{Taylor2013,Russ2015}
\begin{align}
	\bar{H}=\left(
\begin{array}{cccc}
 0 & 0 & \left.t_{l}\right/2 & \left.t_{r}\right/2 \\
 0 & 0 & \sqrt{3}\left.t_{l}\right/2 & -\sqrt{3}\left.t_{r}\right/2 \\
 \left.t_{l}\right/2 & \sqrt{3}\left.t_{l}\right/2 & \Delta +\varepsilon  & 0 \\
 \left.t_{r}\right/2 & -\sqrt{3}\left.t_{r}\right/2 & 0 & \Delta -\varepsilon 
\end{array}
\right)\ko
\label{eq:hubmatrix}
\end{align}
where $t_{l,(r)}$ is the hopping between the left (right) and the center QD, $\varepsilon$ is the energy difference between the outer QDs and $\Delta$ is the effective energy difference between the center QD and the outer QDs in which Coulomb repulsion is included (see Fig.~\ref{fig:TQD}~(a)). In the RX regime ($|\varepsilon|<|\Delta|$ and $t_{l,r}\ll |\Delta \pm \varepsilon| $) the states $ \ket{0}$ and $\ket{1}$ are nearly eigenstates, while the states $ \ket{2} $ with a charge configuration $ (2,0,1) $  and $ \ket{3} $ with a charge configuration $ (1,0,2) $ are only virtually occupied and can be eliminated via a second order Schreiffer-Wolff (SW) transformation yielding a Heisenberg model $H_{\text{Heis}}=J_{l}\boldsymbol{S_{1}}\cdot\boldsymbol{S_{2}} + J_{r}\boldsymbol{S_{2}}\cdot\boldsymbol{S_{3}}$ with the exchange energies $J_{l}\equiv t^{2}_{l}/(\Delta + \varepsilon)$ and $J_{r}\equiv t^{2}_{r}/(\Delta - \varepsilon)$. In its eigenbasis which we consider as our logical qubit space, the Hamiltonian takes the form,
\begin{align}
	H\st{RX} = \frac{\hbar}{2}\omega\st{RX}\sigma_z
	\label{eq:RX_Ham_ULD}
\end{align}
in which $\hbar \omega\st{RX}\equiv \sqrt{(J_{l}+J_{r})^{2} + 3 (J_{l}-J_{r})^{2}}/2 $ is the resonance frequency of the qubit. Periodic driving of the detuning parameters $ \varepsilon $ and $ \Delta $ allows Rabi transitions between the two eigenstates that are, hence, interesting for cavity quantum electrodynamics (cQED).

For the resonator we consider a superconducting strip-line cavity whose resonance frequency is in the GHz-regime and matches the energy splitting of the RX qubit.\cite{Medford2013,Taylor2013,Russ2015} Since the relevant dynamics of the cavity mostly depends on the dynamic of a single electromagnetic mode near the qubit resonance frequency, the Hamiltonian is that of a single mode,\cite{cohen1997}
\begin{align}
	H\st{cav}=\hbar\omega\st{ph}\left(a^\dagger a + \frac{1}{2}\right),
	\label{eq:cavity}
\end{align}
where $ a^\dagger $ ($ a $) is the bosonic creation (annihilation) operator of a photon with frequency $ \omega\st{ph} $. The associated eigenenergies are $ E\st{cav} = \hbar\omega\st{ph}\left(n\st{ph} + 1/2\right) $, where $ n\st{ph} \equiv \braket{a^\dagger a} $ counts the number of photons with the cavity frequency $ \omega\st{ph} $.

For the qubit-cavity interaction, we consider the minimal coupling Hamiltonian in the dipole approximation near the resonance\cite{cohen1997}
\begin{align}
		H_\text{int}=-\frac{e}{m_\text{eff}}\left(\frac{\hbar}{2\epsilon_0\epsilon V\omega\st{ph}}\right)^{1/2}\bo{\epsilon}_{p}\cdot \bo{p}\left(a+a^\dagger\right),
		\label{eq:lightmatter}
\end{align}
where $ \epsilon $~($ \epsilon_0 $) describes the dielectric constant of the material (vacuum), $ V $ is the volume of the cavity, and $ \bo{\epsilon}_{p} $ is the polarization of the photons. In the logical qubit subspace, the dipole matrix element of interest is
\begin{align}
	g_r\equiv -\frac{e}{2m}\left(\frac{\hbar}{2\epsilon_0\epsilon V\omega\st{ph}}\right)^{1/2}\bra{0}\bo{\epsilon}_{p}\cdot\bo{p}\ket{1}
	\label{eq:dipole}
\end{align}
which describes the photon-induced transition between our two qubit states $ \ket{0} $ and $ \ket{1} $.

%%%%%%%%%%%%%%%%%%%%%%%%%%%%%%%%%%%%%%%%%%%%%%%%%%%%%%%%%%%%%%%%%%%%%%%%%%%%%%%%%%%%%%%%%%%%%
\section{Qubit-cavity coupling}
\label{sec:qubit_cavity}
\subsection{Phenomenological approach}
\label{ssec:ham_ideal}

\begin{figure}
		\begin{center}
				\includegraphics[width=1\columnwidth]{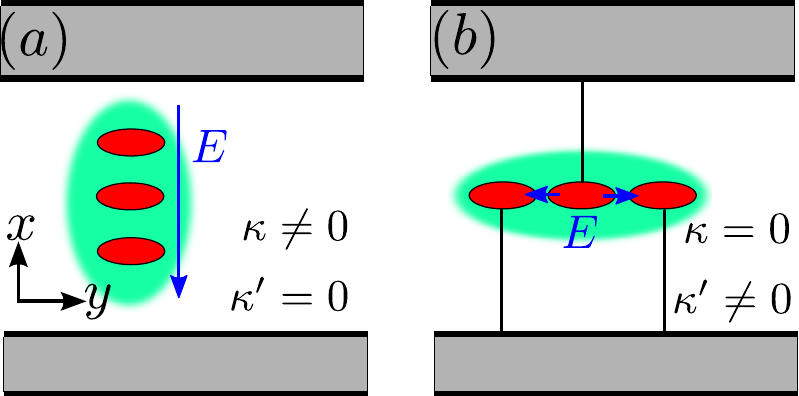}
				\caption{Two possible arrangements of the electric field $E$ (blue arrow) in a cavity. $(a)$ The TQD is arranged parallel to the electric field resulting in a finite coupling $\kappa$ to the parameter $\varepsilon$ and in a vanishing transversal qubit-cavity coupling $\kappa^{\prime}$ to $\Delta$, due to the long wave length of the cavity photons, compared to $a_{l,r}$. $(b)$ The opposite case in which the two outer gates of the QDs are connected to the same potential and the electric field $E$ is aligned from the center QD which results in vanishing qubit-cavity coupling $\kappa$ due to a static $\varepsilon$ and a finite $\kappa^{\prime}$.}
				\label{fig:cavity_zoom}
		\end{center}
\end{figure}

To describe the coupling of a qubit to the cavity, we quantize the two detuning parameters $ \varepsilon,\Delta $. This yields $ \varepsilon \rightarrow \varepsilon + \kappa \left(a + a^\dagger\right) $ and $ \Delta \rightarrow \Delta + \kappa^\prime \left(a + a^\dagger\right) $, where $ \kappa $ and $ \kappa^\prime $ are parameters that include both, the amplitude seen by the qubit for a given polarization and the vacuum amplitude of the electric field, and $ a^\dagger $ ($ a $) is again the creation (annihilation) operator of a photon in the cavity with frequency $ \omega\st{ph} $. Assuming $ \kappa \left\langle a+a^\dagger\right\rangle \ll \varepsilon $ and $ \kappa^{\prime} \left\langle a+a^\dagger\right\rangle \ll \Delta $ we can expand the RX Hamiltonian in terms of $ \kappa \left(a+a^\dagger\right) $ and $ \kappa^{\prime} \left(a+a^\dagger\right) $. As a result, we find\cite{Russ2015}
\begin{align}
	H\st{RX} = \frac{1}{2} \big[ (\hbar\omega\st{RX} + \delta\omega_z)\sigma_z  + \delta\omega_x\sigma_x\big]\ko
	\label{eq:expandH}
\end{align}
represented in the eigenbasis of the unperturbed system with the longitudinal coupling $ \delta\omega_z =  -(J \delta J + 3 j \delta j) \left(a+a^\dagger\right)/\omega\st{RX}$ and the transversal coupling $ \delta\omega_x = \sqrt{3} (J \delta j - j \delta J)\left(a+a^\dagger\right)/\omega\st{RX}$. Here, we used $J\equiv(J_{l}+ J_{r})/2$, $\delta J \equiv \partial_{\varepsilon} J\,\kappa+ \partial_{\Delta} J\,\kappa^{\prime}$, $j\equiv(J_{l}- J_{r})/2$, $\delta j \equiv \partial_{\varepsilon} j\,\kappa+ \partial_{\Delta} j\,\kappa^{\prime}$ and the derivatives $\partial_{\varepsilon} J = \partial_{\Delta} j =(\varepsilon J-\Delta j)/(\Delta^{2}-\varepsilon^{2})$, $\partial_{\varepsilon} j = \partial_{\Delta} J =(\varepsilon j-\Delta J)/(\Delta^{2}-\varepsilon^{2})$. Neglecting higher-order terms of the expansion which correspond to two-photon processes and higher, we obtain, 
\begin{align}
		H\st{JC}=\frac{\hbar\omega_{s}}{2}\sigma_z+g\st{s}\,\sigma_x \left(a+a^\dagger\right)+\hbar\omega\st{ph} \left(a^\dagger a+1/2\right)
		\label{eq:JC_Ham}
\end{align}
with the coupling parameter
\begin{align}
g\st{s} = \sqrt{3} (J \delta j - j \delta J) /\omega\st{RX}
\label{eq:phen}
\end{align}
and a photon-dependent resonance frequency $\omega_{s}\equiv \omega\st{RX} -(J \delta J + 3 j \delta j) \left(a+a^\dagger\right)/\omega\st{RX}$. For certain alignments of the TQD and the cavity, the coupling parameters can be set, e.g., for an alignment as in Fig.~\ref{fig:cavity_zoom}~(a) we expect $\kappa^{\prime}\approx 0$ due to the long wavelengths of the cavity photons while for the alignments in Fig.~\ref{fig:cavity_zoom}~(b) $\kappa$ is negligible. In the case $\kappa^{\prime}=0$, we obtain
\begin{align}
g\st{s} =\sqrt{3}\kappa \frac{\Delta}{\omega\st{RX}} \frac{t_{l}^{2}t_{r}^{2}}{\left(\Delta^{2}-\varepsilon^{2}\right)^{2}}
\label{eq:kappa}
\end{align}
and in the case $\kappa=0$, we find
\begin{align}
g\st{s} =\sqrt{3}\kappa^{\prime} \frac{\varepsilon}{\omega\st{RX}}\frac{t_{l}^{2}t_{r}^{2}}{\left(\Delta^{2}-\varepsilon^{2}\right)^{2}}.
\end{align}

Considering incoherent and broad-band electromagnetic fields, we find the same expression Eq.~(\ref{eq:expandH}) for the RX qubit under the influence of charge noise.\cite{Russ2015} This is not surprising, since in both cases the RX qubit is disturbed by electromagnetic fields. Hence, increasing the qubit-cavity coupling also increases the coupling to charge noise and moving to a sweet spot, where the qubit is robust against this noise,\cite{Taylor2013,Russ2015} we expect a weak qubit-cavity coupling as a trade-off. This phenomenological model does not provide a microscopic description of the coupling parameters $\kappa$ and $\kappa^{\prime}$. We consider a more realistic model for an alignment as in Fig.~\ref{fig:cavity_zoom}~(a) in the following Section~\ref{sec:microscopic}, which will also allow us to estimate $\kappa$.

\subsection{Microscopic theory}
\label{sec:microscopic}
Coupling the qubit states of the RX qubit in the TQD to the cavity via the emission and absorption of a cavity photon requires a strong electric-dipole transition element $g_{r}$ between the two qubit states. We consider an alignment of the TQDs as shown in Fig.~\ref{fig:cavity_zoom}~(a) with the electric field in the cavity pointing in $x$-direction. To find a finite electric-dipole transition in this system, the following conditions have to be fulfilled. Firstly, the TQD has to be coupled by inter-dot exchange interactions, i.e., hopping between neighboring QDs. We find that the matrix element is approximately proportional to the energy splitting between the qubit states ($\simeq t_{l}t_{r}$) which matches with past calculations with a double quantum dot (DQD).\cite{Burkard2006} However, there are still two independent symmetries which have to be broken for a non-vanishing matrix element.

The first symmetry arises from the spin-conserving nature of the electric-dipole transitions and corresponds with inversion symmetry. To distinguish the product states of three electrons in a linearly arranged TQD we use three quantum numbers.~\cite{Hung2014} The first two are the total spin $S$ and its $z$-component $S_{z}$ which are identical for the qubit states, while the third quantum number is the total spin $S_{O}$ of the two electrons in the outer QDs 1 and 3 which distinguishes them, $S_{O}\ket{0}=0$ and $S_{O}\ket{1}=1$, hence, they cannot be transferred into each other by an interaction $O$ with $[S_{O},O]=0$. The influence of the states $\ket{2}$ and $\ket{3}$ with asymmetric charge configuration breaks this symmetry for $J_{l}\neq J_{r}$, e.g., through a gate detuning $\varepsilon\neq 0$.

The second symmetry that needs to be broken is an orbital symmetry between the electrons in the outer QDs. The orbital wave functions of an electron in QD~1 and 3 have identical parity under the assumption that all three QDs have the same lowest orbital confinement potential. To overcome this symmetry either the confinement potentials of the two outer QDs need to be different or the inter-dot distance between the TQDs has to be asymmetric, hence, $a_{l}\neq a_{r}$ where $a_{l,(r)}$ is the inter-dot distance between QD~1 (QD~3) and the center QD (see Fig.~\ref{fig:TQD}~(b)).

\subsubsection{Wave functions}
\label{ssec:OWOa}

To calculate the transition dipole matrix element $ g_{r}\propto\bra{0} \boldsymbol{\epsilon}_{p} \cdot\boldsymbol{p} \ket{1} $ between the qubit states, we need orbital wave functions which are pairwise orthogonal, since the isolated orbital wave functions have a finite overlap (see Fig.~\ref{fig:TQD}~(b)). For orthogonal states, these finite overlaps $S_l = \braketv{\varphi_1}{\varphi_2}$, $ S_r= \braketv{\varphi_3}{\varphi_2}$ and $ S_{13} = \braketv{\varphi_1}{\varphi_3}$ have to be zero which is not the case here. Therefore, we transform the non-orthogonal basis $ \lbrace\ket{\varphi_1},\ket{\varphi_2},\ket{\varphi_3}\rbrace $ into an orthonormal basis of the Wannier orbitals $ \lbrace\ket{\Phi_1},\ket{\Phi_2},\ket{\Phi_3}\rbrace $ which fulfill the orthonormality condition $  \braketv{\Phi_i}{\Phi_j} =\delta_{ij}$ with $ i,j\in\lbrace 1,2,3 \rbrace$. Here, these states are chosen such that they describe the dynamics in the $ i $-th QD and converge for long distances to the isolated electron wave functions, $\lim\limits_{a_l,a_r\rightarrow \infty} \ket{\Phi_i} = \ket{\varphi_i}$. Hence, we obtain in the general case,
\begin{align}
\begin{split}
	\ket{\Phi_1} &= \frac{1}{N_1}\left(\ket{\varphi_1}+ a_1\ket{\varphi_2} +b_1\ket{\varphi_3} \right) \\
	\ket{\Phi_2} &= \frac{1}{N_2}\left(\ket{\varphi_2}+ a_2\ket{\varphi_1} +b_2\ket{\varphi_3} \right)\\
	\ket{\Phi_3} &= \frac{1}{N_3}\left(\ket{\varphi_3}+ a_3\ket{\varphi_2} +b_3\ket{\varphi_1} \right),
\end{split}
	\label{eq:WO_gen}
\end{align}
where $ a_i $ and $ b_i $ are bounded, real parameters. However, these Wannier states are not uniquely defined, since the orthogonality condition yields only a linear equation system with three equations, but, with nine independent parameters. Three parameters ($ N_1,N_2,N_3 $) can be eliminated immediately by the normalization condition,
\begin{align}
\begin{split}
	N_1 &= \sqrt{1+ 2a_1\, S_l + 2b_1\,S_{13}+2a_1\,b_1\,S_r+a_1^2+b_1^2},\\
	N_2 &= \sqrt{1+ 2a_2\, S_l + 2b_2\,S_r+2a_2\,b_2\,S_{13}+a_2^2+b_2^2},\\
	N_3 &= \sqrt{1+ 2a_3\, S_r + 2b_3\,S_{13}+2a_3\,b_3\,S_l+a_3^2+b_3^2}.
\end{split}
	\label{eq:WO_norm}
\end{align}
In a long distance approximation, we neglect the overlap between QD 1 and QD 3 and set $ S_{13}=b_1=b_3=0 $. This reduces the number of free parameters by two and yields simpler expressions for the normalization parameter, $ N_1=\sqrt{1+2a_{1}\, S_{l} + a_{1}^{2}}$, $N_3=\sqrt{1+2a_{3}\, S_{r} + a_{3}^{2}} $ and $ N_2=\sqrt{1+ 2a_2\, S_l + 2b_2\,S_r+a_2^2+b_2^2} $. The last parameter we adapte from the condition of maximally localized Wannier orbitals where we minimize the localization functional\cite{Marzari2012} $\mathcal{F}=\sum_{i=1}^3 \left( \bra{\Phi_i}\hat{x}^2\ket{\Phi_i} - \bra{\Phi_i}\hat{x}\ket{\Phi_i}^{2}\right) $ (see Appendix~\ref{app:mini}). We found at lowest order a simple behavior for the  parameters $ a_1 $ and $ a_3 $, in particular, $ a_1= \xi_a S_l + \mathcal{O}(S_r,S_l^2) $ and $ a_3= \xi_d S_r + \mathcal{O}(S_l,S_r^2) $ and $ \xi_a =\xi_d $ which yields the final condition $a_1/a_3 =S_l/ S_r $. As a result, we obtain analytical expressions for the parameters, $a_1= -2S_l$, $a_2=S_l/(1-2\,S_l^2-2\,S_r^2)$, $b_2=S_r/(1-2\,S_l^2-2\,S_r^2)$ and  $a_3= -2S_r$ within the scope of the approximation. A detailed parameter discussion can be found in Appendix \ref{app:limit}.

%%%%%%%%%%%%%%%%%%%%%%%%%%%%%%%%%%%%%%%%%%%%%%%%%%%%%%%%%%%%%%%

\subsubsection{Transition dipole matrix elements}
\label{ssec:trans_elements}

Having found pair-wise orthonormal wave functions in position space we first calculate the matrix elements of the position operator $ \hat{x} $ in the one-particle basis which we use to calculate $ \bra{0}p_{x}\ket{1} $ and finally $g_{r}$. Therefore, we express the position operator $ \hat{x} $ in the orthonormal basis of the Wannier functions $\lbrace \ket{\Phi_1},\ket{\Phi_2},\ket{\Phi_3}\rbrace$ each combined with one of the two spin states $\ket{\downarrow},\ket{\uparrow} $ in order to describe the electron spin dynamics. As a result, we obtain
\begin{align}
\begin{split}
	\hat{x}=&\sum\limits_{i,j=1}^{3}\sum\limits_{\sigma=\uparrow\downarrow} x_{ij} c_{i,\sigma}^{\dagger}c_{j,\sigma},
	\end{split}
	\label{eq:positionoperator}
\end{align}
where $ x_{ij}= \bra{\Phi_{i}}\hat{x}\ket{\Phi_j} $ with $ i,j\in \lbrace 1,2,3\rbrace $ and the operator $ c_{1,\uparrow}^\dagger $ ($c_{1,\uparrow}$) in Eq.~\eqref{eq:logicalspace2} creates (annihilates) an electron in the orthonormalized Wannier orbital with spin $\sigma\in\lbrace\uparrow\downarrow\rbrace$. In the next step, we construct the qubit states $ \ket{0},\ket{1} $ and the asymmetric states $ \ket{2},\ket{3} $ in terms of the spin Wannier states $ \ket{\Phi_{i,\sigma}} \equiv\ket{\Phi_{i}}\ket{\sigma}$. Keeping this in mind, we now express the position operator $ \hat{x} $ in the $ \lbrace\ket{0},\ket{1},\ket{2},\ket{3}\rbrace $ basis as
\begin{align}
\hat{x}=\left(
	\begin{array}{cccc}
	 \sum\limits_{i=1}^{3}x_{ii} & 0 & \frac{1}{\sqrt{2}}x_{12} & \frac{1}{\sqrt{2}}x_{32} \\
	 0 & \sum\limits_{i=1}^{3}x_{ii} & \sqrt{\frac{3}{2}} x_{12} & -\sqrt{\frac{3}{2}} x_{32} \\
	 \frac{1}{\sqrt{2}}x_{21} & \sqrt{\frac{3}{2}} x_{21} & 2 x_{11}+x_{33} & -x_{31} \\
	 \frac{1}{\sqrt{2}}x_{23} & -\sqrt{\frac{3}{2}} x_{23} & -x_{13} & x_{11}+2 x_{33} \\
	\end{array}
	\right).
	\label{eq:positionoperator2}
\end{align}

We obtain the elements of the momentum operator in the same basis through the relation $ \bo{p} = -\frac{\I\, m}{\hbar}\left[H\st{Hub},\bo{x}\right]$.~\cite{cohen1997} Here, $ H\st{Hub} $ is the full Hamiltonian in this basis given in Eq.~\eqref{eq:hubmatrix} and the square brackets denote the commutator. An analytical expression can be obtained and is shown in Appendix \ref{eq:app_momentum_full}. The simplified expression in Eq.~\eqref{eq:app_momentum_simp} can be obtained considering real matrix elements, $ x_{ij} = x_{ji} $ with $ i,j\in\lbrace 1,2,3\rbrace $.

\subsubsection{Interaction Hamiltonian}
\label{ssec:Inter_Ham_Real}

We now have almost all the tools for calculating the qubit-cavity coupling $ g\st{r} $ in $H_{\text{int}}$. However, due to the Schrieffer-Wolff transformation the qubit states in the expression for the RX Hamiltonian in Eq.~\eqref{eq:RX_Ham_ULD} are not exactly the states defined in Eq.~\eqref{eq:logicalspace2}, in particular they have a small contribution from the asymmetric states $ \ket{2} $ and $ \ket{3} $. Therefore we have to shift into the qubit basis by the same transformation, hence, compute $ \tilde{\bo{p}} = \E^{S}\bo{p}\E^{-S} \approx \bo{p} - \left[\bo{p},S\right] $, where $ S $ is the SW transformation matrix.\cite{Taylor2013,Russ2015} Considering only the matrix elements in the logical subspace of the RX qubit $ \lbrace\ket{0},\ket{1}\rbrace $ and neglecting all other elements, we obtain as a result (see Appendix~\ref{app:matrix_mom})
\begin{align}
\begin{split}
		\tilde{p}_x=& \frac{\sqrt{3} m}{\hbar}\,\frac{t_lt_r\varepsilon}{\Delta^2 -\varepsilon^2}\text{Re}\left(x_{13}\right)\sigma_y\\
		&+\frac{m}{\hbar}\Big[\frac{t_{l}t_{r}\varepsilon}{\Delta^{2}-\varepsilon^{2}}\text{Im}\left(x_{13}\right)\\ &+\sqrt{2}t_{r}\text{Im}(x_{23})-\sqrt{2}t_{l}\text{Im}(x_{12}) \Big]\sigma_z\\
		 &-\frac{\sqrt{6} m}{\hbar} \left[ t_{l} \text{Im}(x_{12}) +t_{r}\text{Im}(x_{23})\right]\sigma_{x},
		\end{split}
		\label{eq:matrix_elements}
\end{align}
where irrelevant terms proportional to the identity operator in the qubit space have been omitted.
\begin{figure}[t]
	\begin{center}
		\includegraphics[width=1\columnwidth]{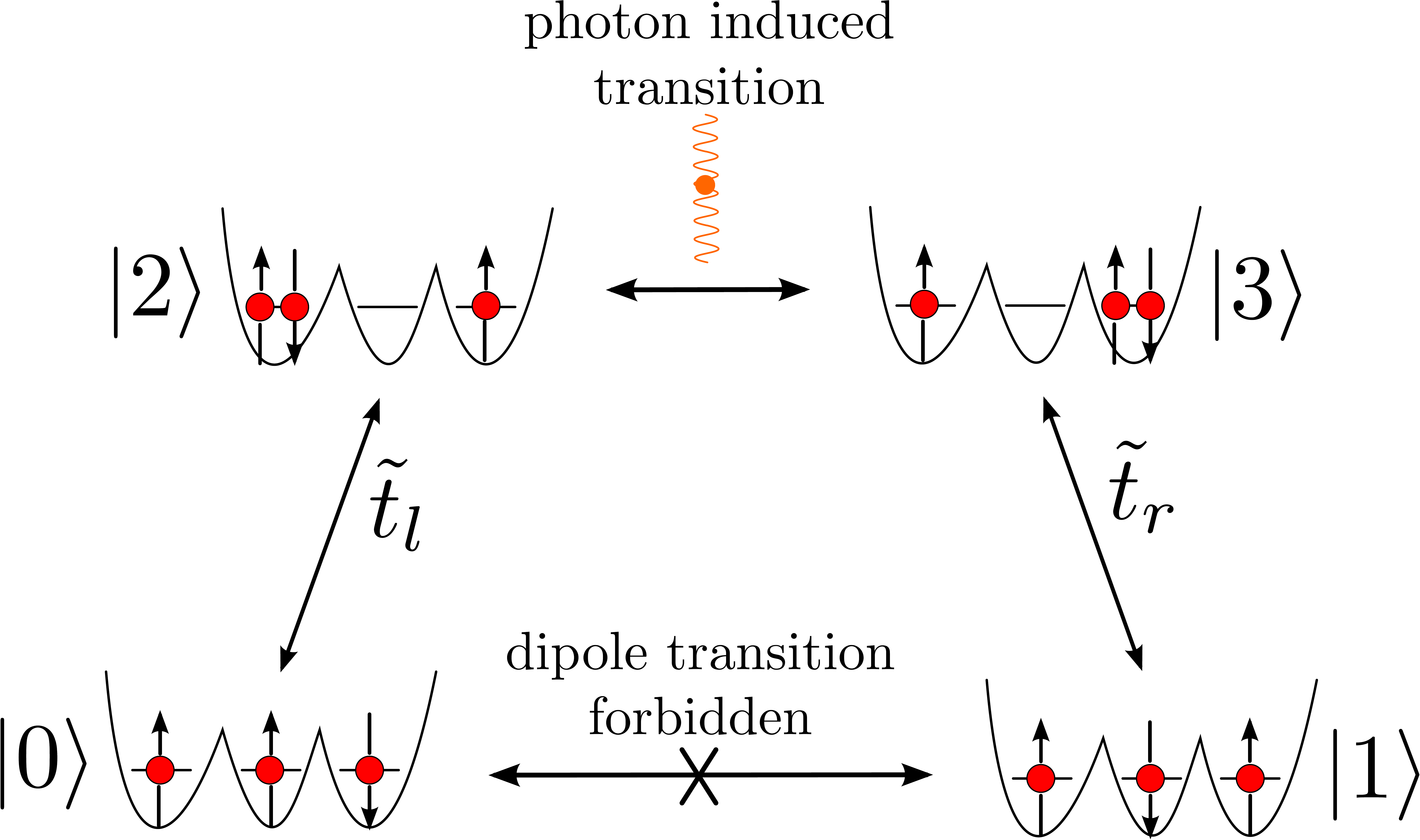}
		\caption{Schematic illustration of the qubit-cavity coupling mechanism. Photons induce dipole transitions between the virtually coupled states $ \ket{2} $ and $ \ket{3} $ (top line) which have an asymmetric charge configuration due to the double occupation of one QD. These two asymmetric states couple weakly through some hopping matrix elements, here denoted as $ \tilde{t}_{l,r} $, to the qubit states. The arrows in $\ket{0}$ and $\ket{1}$ show only one of the possible spin configurations.}
		\label{fig:dipole_coupling}
	\end{center}
\end{figure}

Comparing this result for $\tilde{p}_{x}$ including non-vanishing matrix elements with the one for the bare momentum operator $\hat{p}_{x}$ in the top-left corner of Eq.~\eqref{eq:app_momentum_full}, one finds that in the latter, only the imaginary part of the single electron matrix elements appears, e.g., $\text{Im}(x_{23}) $, whereas Eq.~\eqref{eq:matrix_elements} also contains the real part of $x_{13}$. Hence, we conclude that the physical mechanism of the coupling between the qubit states of the RX qubit is due to the coupling to the asymmetric states $ \ket{2} $ and $ \ket{3} $, since in our case $x_{ij}\in \mathbb{R}$, with $i,j\in\lbrace 1,2,3\rbrace$ (see Appendix \ref{app:calc_coup}). In particular, the electromagnetic field of the cavity induces electronic transitions, where one electron from a doubly occupied QD is transferred to the singly occupied QD and not the empty one, hence, between the states $ \ket{2} $ and $ \ket{3} $ (see Fig.~\ref{fig:dipole_coupling}). In this case, we find 
\begin{align}
	\tilde{p}_x=& \frac{\sqrt{3} m}{\hbar}\,\frac{t_lt_r\varepsilon}{\Delta^2 -\varepsilon^2}x_{13}\sigma_y.
\end{align}It should be noted at this point that a non-perfect alignment of the TQD or electric field in $x$-direction, e.g. a tilting angle in the $xy$-plane, gives rise to an imaginary contribution to the dipole transition matrix elements, hence, allowing for other electronic transitions (see Eq.~\eqref{eq:matrix_elements}). However, this contribution is small if the projection on the $y$-axis, i.e., the tilt angle, is small.
\begin{figure}[t]
	\begin{center}
		\includegraphics[width=1\columnwidth]{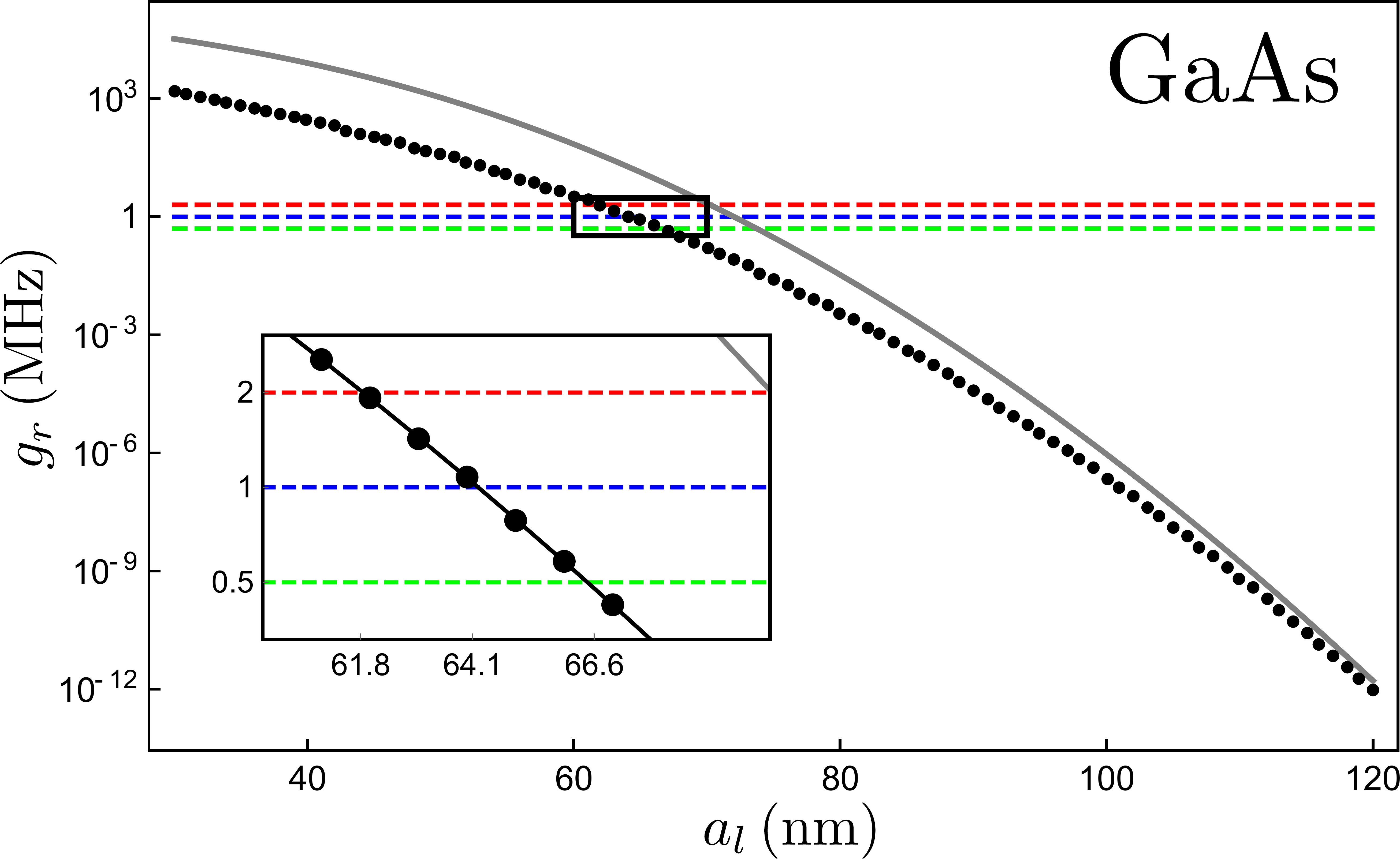}\\
		\vspace{10pt}
		\includegraphics[width=1\columnwidth]{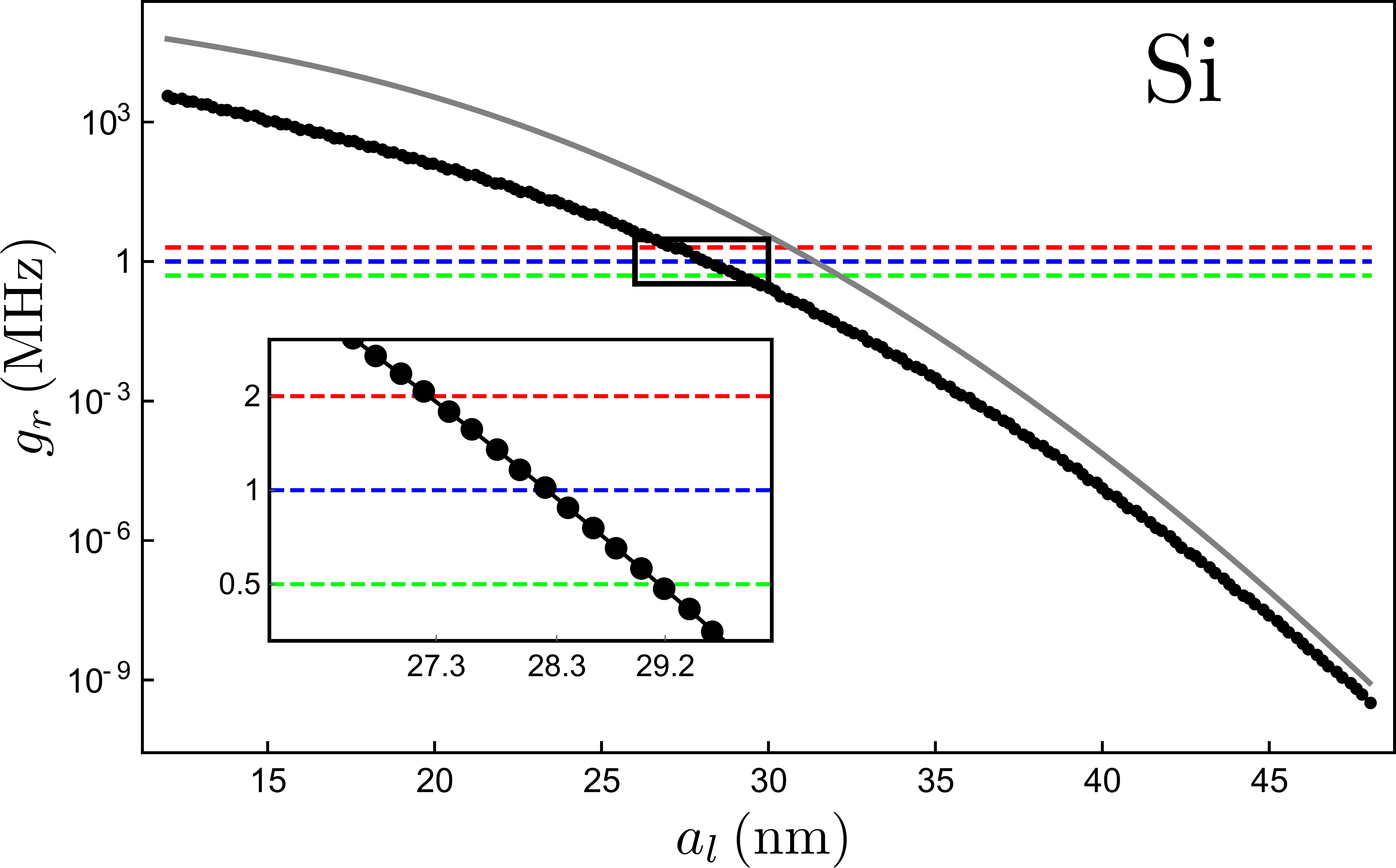}
		\caption{Qubit-cavity coupling $ g_r $ as a function of the inter-dot distance $ a_l $ for (top) GaAs and (bottom) Si. The parameters are chosen as $ a_r=1.1\,a_l $, $ \varepsilon = \unit[0.2]{meV} $, $ \Delta = \unit[0.3]{meV} $, $ B=\unit[310]{mT}$, $ \omega_{0}=\unit[3.1]{\mu eV} $ and $ \omega\st{ph}=\omega\st{RX} \approx\unit[19.54]{\mu eV} $\cite{Russ2015}. We use hopping parameters $ t_l $ and $ t_r $ from numerical calculations\cite{Milkojevic2015} for the black dots, while we consider the expression for a DQD $ t_{l,r} \simeq \hbar  \omega\st{QD}\left[\left(a_{l,r}/a\st{0}\right)^2+\left(\omega_{0}/\omega\st{QD}\right)\right]S_{l,r}/2\left(1-S_{l,r}^2\right) $ for the gray curve.\cite{Burkard1999} Here, $ a\st{0}=\sqrt{\hbar/m\,\omega\st{0}}$ is the confinement radius with $\hbar \omega\st{QD} $ being the confinement energy of the QDs. Typical values are for a GaAs QD $\omega_{\text{QD,GaAs}}=\unit[3.1]{meV}$ and for a silicon QD $\omega_{\text{QD,Si}}=\unit[5.9]{meV}$ due to their different effective masses, $m\st{GaAs}=0.067m_{0}$ and $m\st{Si}=0.191m_{0}$, with $m_{0}$ being the free electron mass. For the volume $ V = L\,d_1\,d_2 $ in the vacuum amplitude of the electric field we set the length $ L=\unit[1]{mm} $, the width $ d_1 = \unit[1]{\mu m} $ and the height $ d_2 = \unit[100]{nm} $ of the stripline cavity. The red, blue and green dashed lines indicate three appropriate values for quantum gate operations.}
		\label{fig:couplingstrength}
	\end{center}
\end{figure}

As a last step, we directly compute the single particle matrix elements in order to find the qubit-cavity coupling $ g_r $. The calculation of the associated matrix element $ x_{13} $ with the help of the orthonormalized Wannier functions $ \ket{\Phi_i} $ yields
\begin{align}
		\tilde{p}_x=-\frac{\sqrt{3}\,m}{\hbar}\,\frac{t_lt_r\varepsilon}{\Delta^2 -\varepsilon^2}\tilde{a}\st{rel} \left(a_l-a_r\right) S_l S_r  \sigma_y\ko
		\label{eq:mom_ULD}
\end{align}
where $\tilde{a}\st{rel}$ is close to one.
An explicit calculation can be found in Appendix \ref{app:calc_coup}. Substituting the resulting momentum operator into Eq.~\eqref{eq:lightmatter} we obtain for the interaction Hamiltonian
\begin{align}
		H_\text{int}=g\st{r}\,\sigma_y \left(a+a^\dagger\right)
		\label{eq:JCy}
\end{align}
with the effective coupling strength
\begin{align}
		g\st{r}=-\frac{\sqrt{3}\,e\,E_0}{2\hbar\omega\st{ph}}\,\frac{t_lt_r\varepsilon}{\Delta^2 -\varepsilon^2}\tilde{a}\st{rel}\left(a_l-a_r\right)S_lS_r
		\label{eq:couplingstrength}
\end{align}
and the vacuum amplitude of the electromagnetic field in the cavity $ E_0\equiv \left(\hbar\omega\st{ph}/2\epsilon_0\epsilon V \right)^{1/2}$. Identifying $g_{r}=g_{s}$ from Eq.~\eqref{eq:kappa}, we find 
\begin{align}
	\kappa = E_{0}e\frac{a_{l}-a_{r}}{2}\,S_{l}\,S_{r}\frac{\omega\st{RX}}{\omega\st{ph}}\frac{\varepsilon}{\Delta}\frac{\Delta^{2}-\varepsilon^{2}}{t_{l}t_{r}}.
\end{align}

On closer examination, the long distance behavior of $ g_r $ is mostly determined by the overlap parameters $ S_{l,r}\propto \E^{-a^2_{l,r}/4a_{S}^{2}} $ from Eq.~\eqref{eq:overlaps}, while the short distance is mostly determined by the hopping parameters. However, we note that from a realistic point of view the shortest distance in Fig.~\ref{fig:couplingstrength} is given by the size of the QDs which is far outside the scope of our approximation, $a_{l,r}>\unit[40]{nm} $ for GaAs and $a_{l,r}>\unit[18]{nm} $ for Si (see Fig.~\ref{fig:limit} in Appendix~\ref{app:limitWO}). Additionally, only $a_l-a_r\neq 0 $ leads to $ g_r\neq 0 $, since, otherwise, the parities of $\ket{\Phi_{1}}$ and $\ket{\Phi_{3}}$ are identical. In Fig.~\ref{fig:couplingstrength} we have plotted the calculated coupling strength $ g_r $ as a function of the inter-dot distance  $ a_l $ for fixed $a_{r}/a_{l}=1.1$. Both $a_{l}\neq a_{r}$ and $\varepsilon\neq 0$ are required to break orbital and spin symmetries, allowing for $g_{r}\neq 0$. We find that a strong qubit-cavity coupling, e.g., $g_r\approx \unit[2]{MHz}$ (red dashed line in Fig.~\ref{fig:couplingstrength}) becomes accessible for an inter-dot distance $a_l \simeq \unit[60]{nm}$, which is inside the scope of our approximation.

%%%%%%%%%%%%%%%%%%%%%%%%%%%%%%%%%%%%%%%%%%%%%%%%%%%%%%%%%%%%%%%

\section{Implementation of long distance interaction}
\label{sec:ULDinter}
The coupling of the RX qubit to the electromagnetic field of a surrounding cavity enables the coherent transfer of information between the qubit system and the cavity. Therefore, if two RX qubits are coupled to the same cavity, one can transfer information between them via the electromagnetic field. The distance of this transfer is limited only by the extension of the cavity. Inserting the interaction Hamiltonian Eq.~\eqref{eq:JCy} into the two-qubit Hamiltonian Eq.~\eqref{eq:Htotal}, we obtain
\begin{align}
		H\st{RW}=\sum\limits_{i=1}^{2}\left[\frac{\hbar\omega_i}{2}\sigma_z +\I\, g_i \left(\sigma_+a-\sigma_-a^\dagger\right)\right] + \hbar\omega\st{ph}a^\dagger a
		\label{eq:rotatingwave}
\end{align}
in the rotating-wave approximation, where we neglect the counter-rotating terms due to $g_r\ll \omega\st{RX}\approx\omega\st{ph}$. Here $ \sigma_\pm\equiv\left(\sigma_x\pm\I\sigma_y\right)/2$ are the ladder operators and $\omega_i$ are the resonance frequencies of the $i$-th RX qubit with $i\in\lbrace 1,2\rbrace$. Hence, we can eliminate the cavity mode by a second order SW transformation.\cite{Burkard2006,Imamoglu1999} As a result, we obtain
\begin{align}
		H\st{tot}=H\st{1Q}+H\st{int} = \sum\limits_{i=1}^{2}\frac{\epsilon\st{eff}}{2}\sigma_z^i+g\st{eff}\left(\sigma_+^1\sigma_-^2+\sigma_-^1\sigma_+^2\right)
		\label{eq:ULD_SW}
\end{align}
with the Stark-shifted energy $ \epsilon\st{eff}= \hbar\omega_i+g_i^2\left(a^\dagger a+1/2\right)/\left[2\hbar\left(\omega_i -\omega\st{ph}\right)\right]$ and an effective two-qubit coupling parameter $ g\st{eff}=g_1g_2\left\lbrace 1/\left[\hbar\left(\omega_1 -\omega\st{ph}\right)\right]\right.+\left.1/\left[\hbar\left(\omega_2 -\omega\st{ph}\right)\right]\right\rbrace $, where $ g_i $ is the coupling strength between the $ i $-th RX qubit and the cavity. A review of the calculation can be found in Appendix \ref{app:SW2}.
\begin{figure}[t]
	\begin{center}
		\includegraphics[width=1\columnwidth]{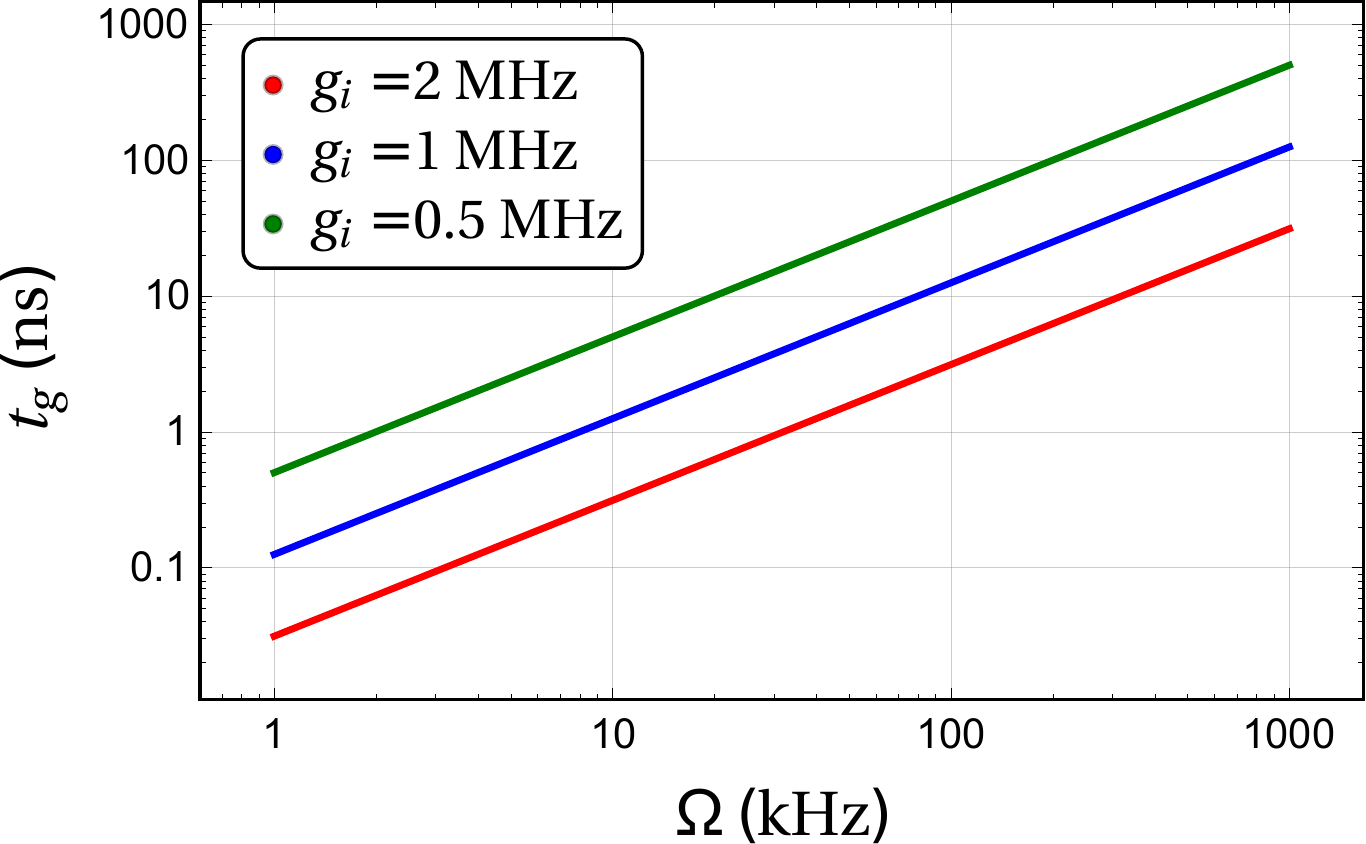}
		\caption{Doubly logarithmic plot of the gate time $ t_g $ of the photon mediated long-distance interaction as a function of the detuning $\Omega= \left|\omega\st{ph} -\omega_{i} \right|/2\pi$. The red, blue and green lines relate to the correspondingly labeled qubit-cavity coupling frequencies $ g_i/(2\pi \hbar) $ in Fig.~\ref{fig:couplingstrength}. For simplicity, we assume that the two coupling frequencies $f_{r}\equiv g_{1,2} /(2\pi \hbar)$ and resonance frequencies $\omega_{1,2}$ each are identical which leads to $t_{g}=\Omega/8f_{r}^{2}$. }
		\label{fig:tuniversal}
	\end{center}
\end{figure}

The first expression $ H\st{1Q} $ in Eq.~\eqref{eq:ULD_SW} yields control over rotations around the z-axis of the qubits, while the interaction part $H\st{int} = g\st{eff}\left(\sigma_+^1\sigma_-^2+\sigma_-^1\sigma_+^2\right) $ leads to an universal two-qubit gate. We obtain for the corresponding time evolution,
\begin{align}
\begin{split}
	U(t,t_0) &= \exp\left(-\I t H\st{int}/\hbar\right)\\
 &=\left(
		\begin{array}{cccc}
		 1 & 0 & 0 & 0 \\
		 0 & \cos\left(g\st{eff}\, t/\hbar\right) & \I \sin\left(g\st{eff}\, t/\hbar\right) & 0 \\
		 0 & \I \sin \left(g\st{eff}\, t/\hbar\right) & \cos \left(g\st{eff}\, t/\hbar\right)  & 0 \\
		 0 & 0 & 0 & 1
		\end{array}
		\right)\,
		\end{split}
\end{align}
which after the time $ t_g=\hbar\pi /\left(2g\st{eff}\right) $ yields the universal i\textsc{swap} gate.\cite{Tanamota2008} In summary, we obtain the i\textsc{swap}-gate with the following four steps:
\begin{enumerate}
\item Prepare both RX qubits to be off-resonant from the cavity frequency, $\omega_{i}\neq\omega\st{ph}$.
\item Detune both RX qubits to be resonant to the cavity frequency, $\omega_{i}\simeq\omega\st{ph}$.
\item Wait for the time $t_g=\hbar\pi /\left(2g\st{eff}\right)$.
\item Detune both RX qubits to be off-resonant from the cavity frequency, $\omega_{i}\neq\omega\st{ph}$.
\end{enumerate}
Together with arbitrary single-qubit gates this allows for universal quantum computation over distances on the order of a the extension of the cavity. Fig.~\ref{fig:tuniversal} shows the pulsing time $t_g$, as a function of the detuning $\Omega_i\equiv \left|\omega\st{ph} -\omega_i \right|/2\pi$ with $ i\in\lbrace 1,2\rbrace $ for specific values of the qubit-cavity couplings $ g_i $. In order to obtain short gating times, we require either a small detuning $\Omega_{1,2} $ or a strong qubit-cavity coupling $ g_i $. We find gate times $t_{g}\approx\unit[1]{ns}$ for $ \Omega\approx \unit[10]{kHz} $ and $ g_i = \unit[1]{MHz} $ which allows for up to $10^3$ gate operations if the qubit decoherence time amounts to $ T_\varphi \approx \unit[1]{\mu s} $.\cite{Medford2013,Taylor2013,Russ2015}

\section{Conclusion and Outlook}
 \label{sec:conclusion_ULD}
In this paper, we have proposed and analyzed an implementation of a long-distance coupling of two RX qubits. We showed that such a coupling can be achieved by placing two RX qubits into a high finesse cavity (Q-factor exceeds $10^{5}$) with the condition that the resonance frequency of the cavity matches the energy separation of the qubit states. By taking into account the wave functions associated with the RX qubit, we have obtained a realistic description of the qubit-cavity interaction which yields a microscopic mechanism for coupling the qubit to the cavity, namely, by a transition of an electron between the outer QDs. In particular, the cavity causes a transition between states $\ket{2}$ and $\ket{3}$ which are coupled to the qubit states by electronic hopping elements.

For the description of the wave functions and in order to estimate the dipole transition matrix element between the qubit states we have used the method of orthonormalized Wannier orbitals to construct the electron wave functions, by taking only a small, finite overlap between the original electron wave functions into consideration. As a result, we obtain the qubit-cavity coupling strength as a function of the inter-dot distances. Realistic parameter settings yield values on the order of $\sim\unit{MHz}$ depending on the chosen distance. Combining these elements enable a coupling between two RX qubits in the same cavity with gate times $t_g$ on the order of $\sim\unit{ns}$ depending on the detuning between the resonance frequency between the cavity and the RX qubits and the qubit-cavity couplings. Realistic parameter choices allow $10^{3}$ operations in the qubit coherence time.

Since the qubit-cavity coupling mechanism is based on the transitions between the asymmetric states $\ket{2}$ and $\ket{3}$ which are only virtually occupied in the RX regime, the asymmetric resonant exchange (ARX) qubit\cite{Russ2015} should have a strong qubit-cavity coupling, since in this implementation the asymmetric states are occupied most of the time. However, this consideration requires a much deeper understanding of the associated wave functions. Nevertheless, we encourage the investigation of these aspects in future studies, since they may strongly increase the qubit-cavity coupling strength.

This far, we have considered in our analysis only a spin-degree of freedom which is appropriate for GaAs. However, silicon has an additional six-fold degeneracy (two-fold in typical QDs) of the ground state, the so-called valley degree of freedom, due to local inequivalent minima (maxima) in the conduction (valence) band.\cite{Zwanenburg2013} This leads to a more complex structure of the single-electron wave functions.\cite{Culcer2009,Culcer2010} In this paper, we considered a non-degenerate ground state (strong valley splitting) with the same valley ground state in all QDs. In future studies, the valley degeneracy can be included, since this additional degree of freedom could possibly be helpful, e.g., serving as additional qubits.\cite{Rohling2012,Rohling2014}

In our analysis, we assumed a harmonic confinement potential of the QDs. Small anharmonicities can give rise to a more complex structure of the wave-functions than the one considered here. In future studies, these modified wave functions can be included to acquire more accurate predictions which will hopefully encourage future experimental implementations of such a setup for long distance interaction.

\section*{Acknowledgments}
We thank Marko Milivojevi\'{c} for helpful discussions and the supply of numerical values for the tunneling matrix elements. We acknowledge funding from ARO through Grant No. W911NF-15-1-0149.

\appendix

\section{Electron wave functions}
\label{app:electron_WF}

Here, we calculate the single electron wave-functions of an isolated electron in a QD defined in a 2D electron system with a perpendicular external magnetic field $ \bo{B} $ applied. We assume that the potential of the quantum dot is approximately harmonic. Hence, the corresponding wave functions of the ground state are given by the following expression,\cite{Burkard1999}
\begin{align}
	\varphi(x,y)=\sqrt{\frac{m\,\omega\st{QD}}{\pi\hbar}}\E^{-m\,\omega\st{QD}\left(x^2+y^2\right)/2\hbar}.
	\label{eq:Gaussian_WF2}
\end{align}
For $B=0$, $ \hbar\omega\st{QD} =\hbar\omega_{0}$ is the harmonic confinement energy of the QD and typically on the order of $ \unit[3]{meV} $ for GaAs which corresponds with a QD diameter $a\st{QD}\approx\unit[20]{nm}$, while the higher confinement energy in Si, e.g., $\unit[6]{meV}$, yields a smaller diameter $ a\st{QD}\approx \unit[9]{nm} $. Due to the magnetic field, $B\approx \unit[310]{mT}$, which is necessary to split off leakage states\cite{Taylor2013}, the wave functions are not properly described by Eq.~\eqref{eq:Gaussian_WF2} and we have to include the influence of the magnetic field. As a result, we obtain the Fock-Darwin states, which are the harmonic states compressed by a factor $ b \equiv \omega\st{QD}/\omega_{0} $, where $ \hbar\omega\st{QD} $ is the confinement potential of the QD under the influence of the magnetic field $ B $ \cite{datta1999}. Here, the modified confinement potential is $ \omega\st{QD}\equiv\sqrt{\omega_{0}^2+\omega_L^2} $ with the Lamor frequency $ \omega_L\equiv e B/2m $. Hence, fixing the origin of the coordinate system to the center QD, we obtain for the single-electron wave-function in the center QD, $\varphi_{2}(x,y)=\varphi(x,y)$.
The wave functions for an electron in the left dot are shifted by $x\rightarrow x+a_l $ and for the right dot by $x\rightarrow x-a_r $, where $ a_l $ ($ a_r $) is the distance from the center dot to the left (right) dot. However, due to the gauge transformation of the magnetic field $\bo{A} = \bo{B}(-y,x\mp a_{l,r})\rightarrow \bo{B}(-y,x)$ they also obtain a phase shift
\begin{align}
\begin{split}
		\varphi_1(x,y)=& \sqrt{\frac{m\,\omega\st{QD}}{\pi\hbar}}\E^{- \I y a_l/2l_B^2}\E^{-m\omega\st{QD}\left[(x+a_l)^2+y^2\right]/2\hbar},\\
		\varphi_3(x,y)=&\sqrt{\frac{m\,\omega\st{QD}}{\pi\hbar}} \E^{\I y a_{r}/2l_B^2}\E^{-m\omega\st{QD}\left[(x-a_r)^2+y^2\right]/2\hbar},
		\end{split}
		\label{eq:FockDarwin}
\end{align}
with the magnetic length $ l_B\equiv\sqrt{\hbar/eB} $. For the overlaps between the left and center QDs, $ S_l\equiv \braketv{\varphi_1}{\varphi_2}$, between the right and center QDs, $ S_r\equiv \braketv{\varphi_3}{\varphi_2}$, and between the left and right QDs, $ S_{13}\equiv \braketv{\varphi_1}{\varphi_3}$, we obtain
 \begin{align}
 		S_{l,r}&=\exp\left[-a_{l,r}^2/ 4a_{S}^{2}\right],\\
 		S_{13}&=\exp\left[-(a_l+a_r)^2 /4a_{S}^{2}\right],
 		\label{eq:overlaps}
 \end{align}
 with the magnetic field dependent QD Bohr radius of the electron wave functions
 \begin{align}
 a_{S}=\left(\frac{\hbar}{4\, l_b^4\, m\, \omega\st{QD}}+\frac{m \omega\st{QD}}{\hbar}\right)^{-1/2}.
 \end{align}For $B=0$, we have $a_{S}=a\st{0}=\sqrt{\hbar/m\omega\st{0}}$.
Typical values are $a_{S}\approx\unit[9.5]{nm}$ for GaAs and $a_{S}\approx\unit[4]{nm}$ for Si.
Without loss of generality, we find that the overlaps $S_{l,r}$ and $S_{13}$ are always real due to our choice of the wave functions and our assumptions of identical confinement potentials $\omega\st{QD}$ in each QD. 

Usually, the overlap $ S_{13} $ is negligible, e.g., for a symmetric setup, $ a_l=a_r $, we obtain for the overlaps of the neighboring QDs $ S_l=S_r\equiv S $ with $S\ll 1$ and for the overlap between the left and right QD $ S_{13}=S^4 $ and we only have to consider $ S_{13} $ for very small distances between the QDs.

%%%%%%%%%%%%%%%%%%%%%%%%%%%%%%%%%%%%%%%%%%%%%%%%%%%%%%%%%%%%%%%
\section{Orthonormalized Wannier orbitals}
\subsection{Minimizing the localization functional}
\label{app:mini}
\begin{figure}[t]
	\begin{center}
	\includegraphics[width=1.\columnwidth]{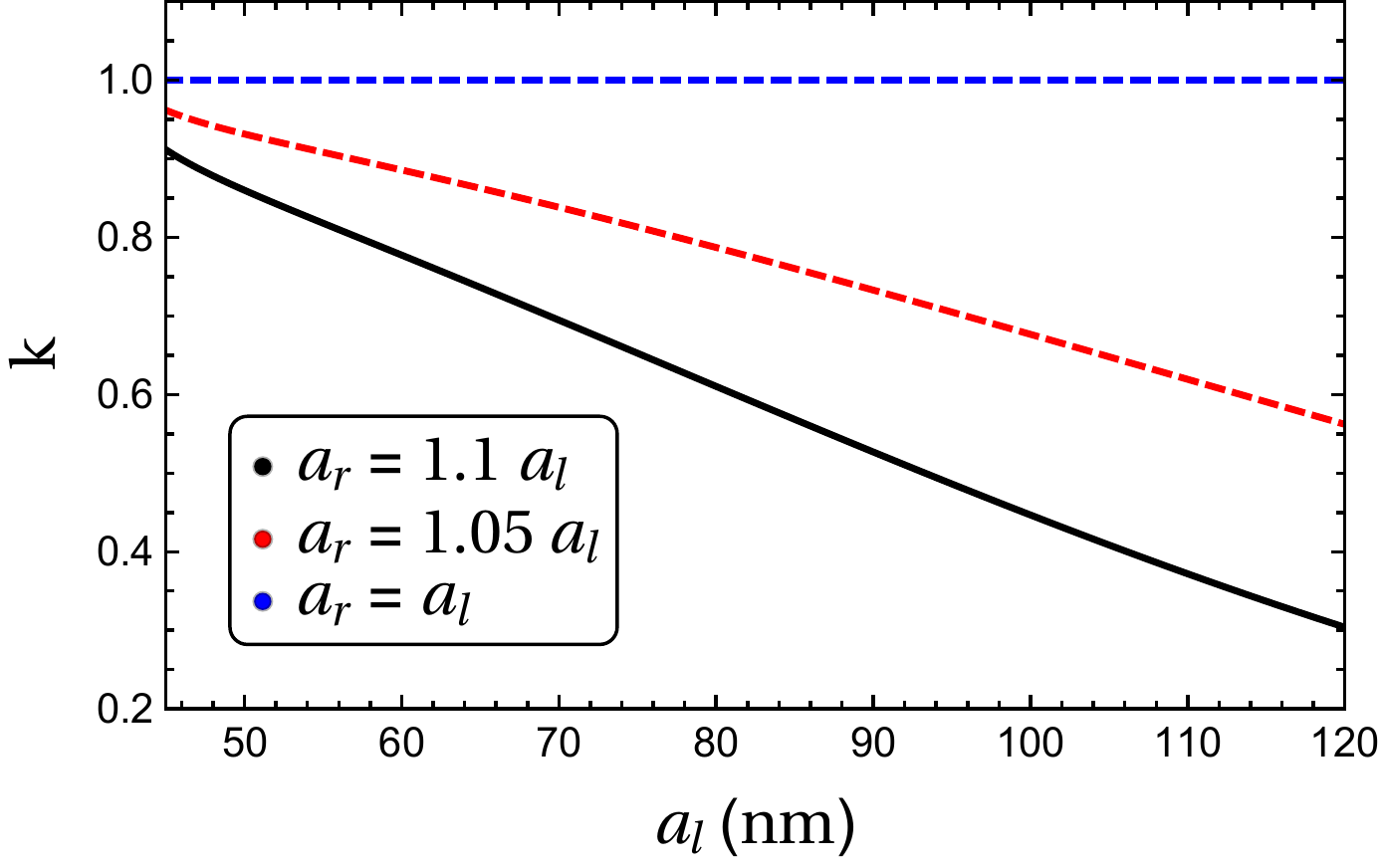}
	\caption{Maximally localized parameter settings for $k$ as a function of the inter-dot distance $a_l$ with different dynamically fixed values for $a_r$. In the symmetric case, $a_l=a_r$, the optimal setting is $k=1$, while for the asymmetric case, $a_l<a_r$, the optimal setting is $0.3\le k\le 1$ depending on the asymmetry and the inter-dot distance. Note that for very long inter-dot distances $k\rightarrow -1$ which corresponds to $\ket{\Phi_{i}} \rightarrow \ket{\varphi_{i}} $.}
	\label{fig:mini}
	\end{center}
\end{figure}
\begin{figure}[t]
	\begin{center}
	\includegraphics[width=1.\columnwidth]{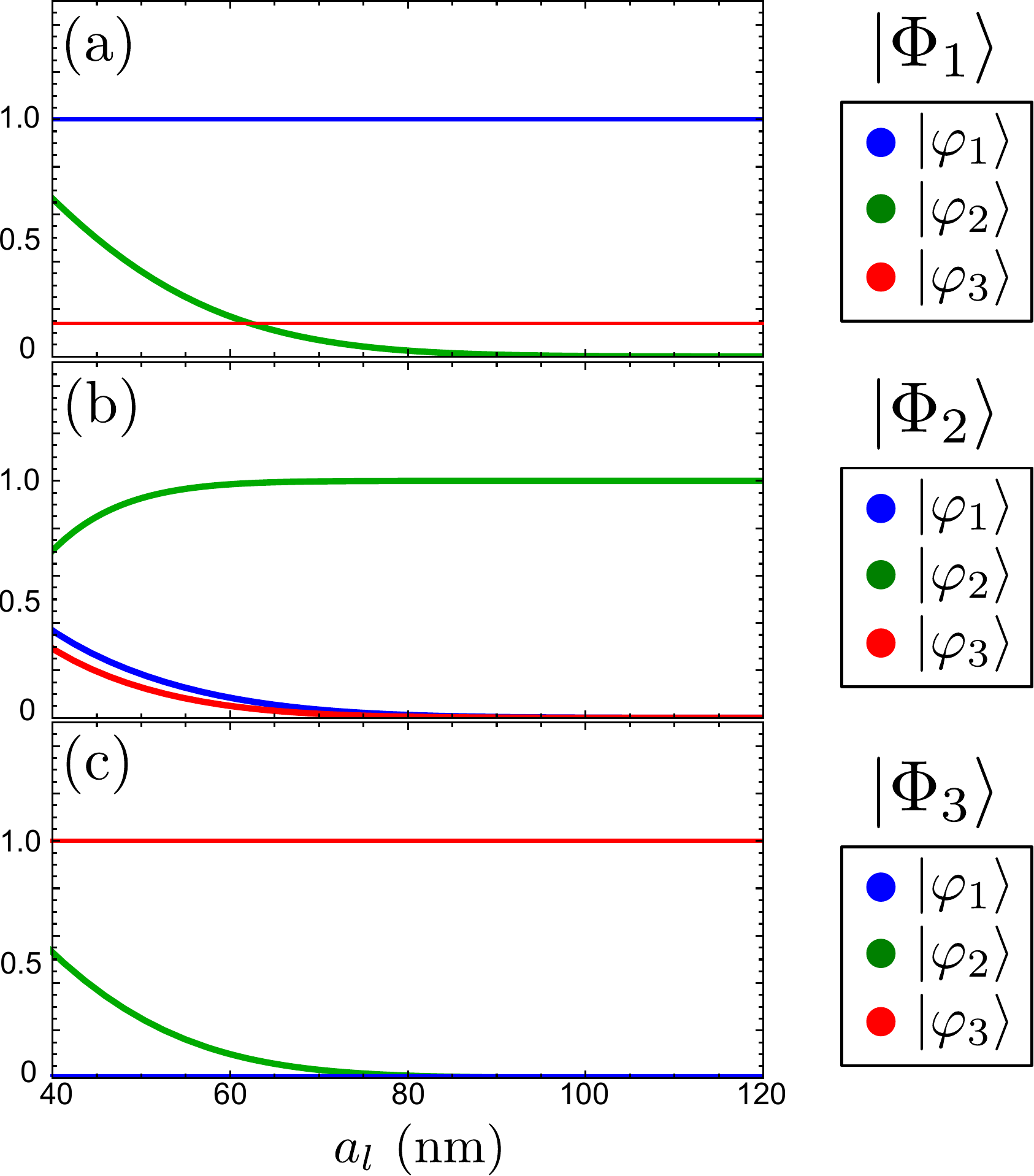}
	\caption{(a)-(c) Constitution of the orthonormalized Wannier orbitals. The colored lines illustrate the overlaps $\left| \braketv{\varphi_i}{\Phi_{j}} \right|$ with $ i,j\in\lbrace 1,2,3\rbrace $ as a function of $ a_l $ with $ a_r = 1.1 a_l$ and $ w\st{QD} = \unit[3.10]{meV} $ in GaAs.}
	\label{fig:overlap}
	\end{center}
\end{figure}
To determine the orthonormalized Wannier orbitals in the long inter-dot distance approximation we use as an ansatz,
\begin{align}
\begin{split}
	\ket{\Phi_1} &= \frac{1}{N_1}\left(\ket{\varphi_1}+ a_1\ket{\varphi_2} \right),\\
	\ket{\Phi_2} &= \frac{1}{N_2}\left(\ket{\varphi_2}+ a_2\ket{\varphi_1} +b_2\ket{\varphi_3} \right),\\
	\ket{\Phi_3} &= \frac{1}{N_3}\left(\ket{\varphi_3}+ a_3\ket{\varphi_2}  \right),
\end{split}
	\label{eq:WO_app}
\end{align}
with parameters that are determined by the orthonormalization condition $\braketv{\Phi_i}{\Phi_j}=\delta_{ij}$ with $i,j\in\lbrace 1,2,3\rbrace$ except for one parameter which we define as $k = a_1 S_r /(a_3 S_l)$. In the next step, we calculate the localization functional\cite{Marzari2012} $\mathcal{F}=\sum_{i=1}^3 \left( \bra{\Phi_i}\hat{x}^2\ket{\Phi_i} - \bra{\Phi_i}\hat{x}\ket{\Phi_i}^{2}\right) $ as a function of $k$ which is straight-forward since the Wannier orbitals are superpositions of Gaussian wave functions. The minimizing conditions, $\partial_k\mathcal{F}=0$, yields two solutions $k\sim\pm1$ in the parameter regime investigated here, $a_l=\unit[45-120]{nm}$ (more details are given in Appendix \ref{app:limit}), where we neglect the negative solution, since it corresponds to a vanishing overlap between the wave functions. Fig.~\ref{fig:mini} shows the best parameter settings from the minimizing condition from numerical calculations. In Fig.~\ref{fig:overlap}, we plot the coefficients of Eq.~\eqref{eq:WO_app} as a function of the inter-dot distance and $k=1$.

\subsection{Limitation of our model}
\label{app:limitWO}
\begin{figure*}[t]
	\begin{center}
	\includegraphics[width=1.\columnwidth]{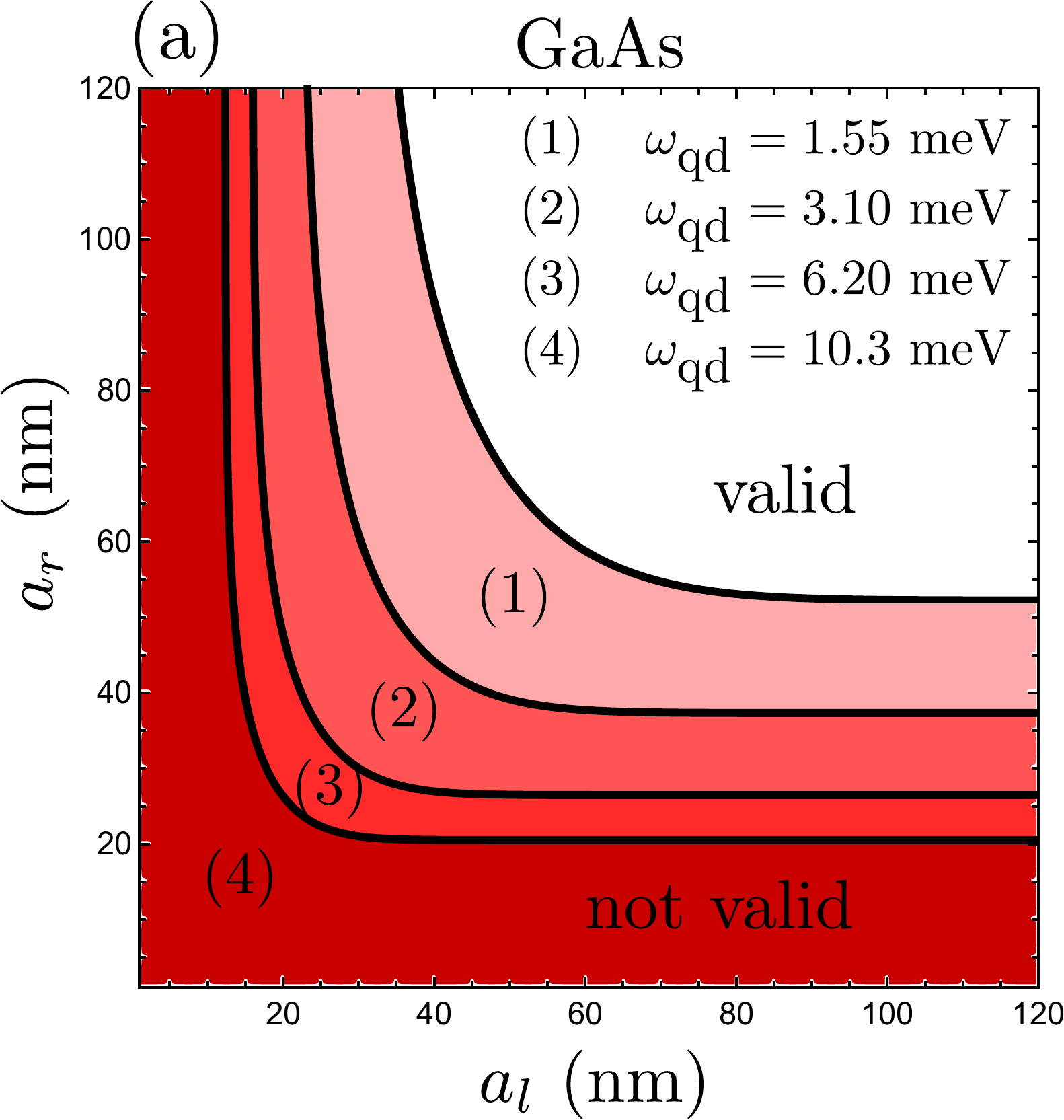}
	\includegraphics[width=0.97\columnwidth]{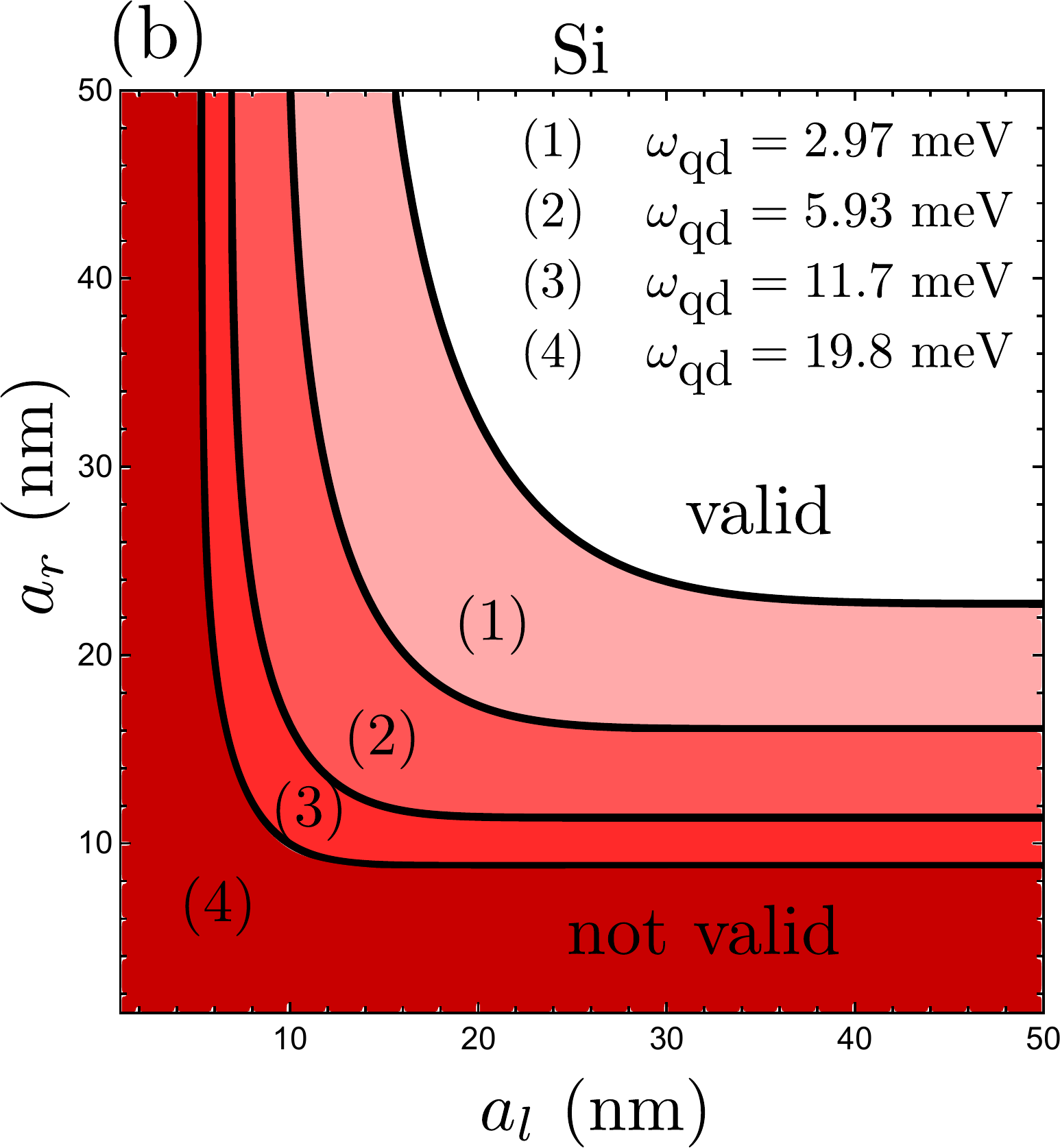}
	\caption{Limitation of the use of orthonormalized Wannier orbitals in (a) a GaAs TQD and (b) a Si TQD in a magnetic field $ B = \unit[310]{mT} $ for different settings of the harmonic potential energies $ \omega\st{QD}$. Here, the area in red shows the case in which the overlap between the QD~1 and QD~3 cannot be neglected and our approximation breaks down.}
	\label{fig:limit}
	\end{center}
\end{figure*}

The limitations of the orthonormalized Wannier orbitals are illustrated in Fig.~\ref{fig:limit} as a function of the inter-dot distances. The Wannier orbital wave functions, $ \ket{\Phi_{i,s}} $, only describe the long inter-dot distance limit. As a condition for the validity of our approach, we use
\begin{align}
		a_{l,r}^2 \gtrsim -2a^{2}_{S} \log\left(\frac{1}{2}-S^2_{r,l}\right).
		\label{eq:limit}
\end{align}Fig.~\ref{fig:limit} shows the resulting limitations for a magnetic field of $ B=\unit[310]{mT} $ for different settings of the confinement energy $ \omega\st{QD}$. Considering $ \hbar\omega\st{QD}=\unit[3.1]{meV} $, we obtain $ a_{l,r}>\unit[40]{nm} $ in GaAs and $ a_{l,r}>\unit[18]{nm} $ in Si.
\label{app:limit}

\section{Matrix elements of the position operator $ \hat{x} $}

For an estimation of the qubit-cavity coupling, the matrix elements of the position operator from Eq.~\eqref{eq:positionoperator} in the $ \lbrace\ket{0},\ket{1},\ket{2},\ket{3}\rbrace $ basis are needed. Therefore, we express all three-electron states in terms of our orthonormalized Wannier functions. In other words, $ c^\dagger_{i,\sigma} $ ($ c_{i,\sigma} $ ) creates (annihilates) an electron in the $ i $-th Wannier orbital with spin $ \sigma $. Hence, we have to calculate the following matrix elements, using Eq.~\eqref{eq:positionoperator},
\begin{align}
	\bra{u}\hat{x}\ket{v} =& \sum\limits_{i,j=1}^{3}\sum\limits_{\sigma=\uparrow,\downarrow} x_{ij} \bra{u}c_{i,\sigma}^{\dagger}c_{j,\sigma} \ket{v} 
\end{align}
with $ u,v\in\lbrace 0,1,2,3\rbrace $. For the sake of brevity, we focus here on one of the two relevant cases $ u=2 $ and $ v=3 $ for the qubit-cavity coupling,
\begin{align}
\begin{split}
\bra{2}\hat{x}\ket{3} =\sum\limits_{i,j=1}^{3}\sum\limits_{\sigma=\uparrow,\downarrow} x_{ij} &\bra{\text{vac}}c_{3,\uparrow} c_{1,\downarrow} c_{1,\uparrow}\\ 
\times&c_{i,\sigma}^{\dagger}c_{j,\sigma}c_{1,\uparrow}^\dagger c_{3,\uparrow}^\dagger c_{3,\downarrow}^\dagger\ket{\text{vac}}.
\end{split}
\end{align}
Inserting $ x $ yields only one non-zero term, since all others are zero, since $ \braketv{\Phi_{i,\sigma}}{\Phi_{j,\sigma^\prime}} = \delta_{ij}\delta_{\sigma\sigma^\prime}  $, hence, we obtain
\begin{align}
	\bra{2}\hat{x}\ket{3} = -x_{13}.
\end{align}
The calculations for the other matrix elements can be done the same way, in particular, $\bra{0}x\ket{1}=0$, since two electrons have to be transferred (or no electrons have to be transferred and a spin has to be flipped). Combing all matrix elements, we find the expression in Eq.~\eqref{eq:positionoperator2}.

\begin{widetext}
\section{Matrix elements of the momentum operator $ \bo{p} $}
\label{app:matrix_mom}
Through the position-momentum relation $ \bo{p} = -\frac{\I\, m}{\hbar}\left[H\st{Hub},\hat{\bo{x}}\right]$ we obtain for the momentum operator in $ x $-direction in the $ \lbrace\ket{0},\ket{1},\ket{2},\ket{3}\rbrace $ basis
\begin{align}
\hat{p} = \left(\begin{array}{cc}
				\bo{A} & \bo{B} \\
				\bo{B}^\dagger & \bo{C}
		\end{array}	\right),
	\label{eq:app_momentum_help}
\end{align}
with the $2\times 2$ blocks

\begin{align}
\begin{split}
	\bo{A} &\equiv \left(\begin{array}{cc}
		\frac{\I m (t_l x_{12}-t_l x_{21}+t_r (x_{32}-x_{23}))}{2 \sqrt{2} \hbar}  &  
		\frac{\I\sqrt{\frac{3}{2}} m (t_l x_{12}-t_l x_{21}+t_r (x_{23}-x_{32}))}{2\hbar}\\
		\frac{\I\sqrt{\frac{3}{2}} m (t_l x_{12}-t_l x_{21}+t_r (x_{23}-x_{32}))}{2\hbar} & 
		\frac{3\I m (t_l x_{12}-t_l x_{21}+t_r (x_{32}-x_{23}))}{2 \sqrt{2} \hbar} \\
		\end{array}\right),\\
	\bo{B} & \equiv -\left(\begin{array}{cc}
		\frac{\I m \left(t_l x_{21}-\sqrt{2} (\Delta +\varepsilon ) x_{12}-t_r x_{13}-t_l x_{22}\right)}{2\hbar} &
		\frac{-\I m \left(t_r x_{22}+t_l x_{31}+\sqrt{2} (\Delta -\varepsilon ) x_{32}-t_r x_{33}\right)}{2\hbar} \\
		\frac{\I \sqrt{3} m \left(t_l x_{11}-\sqrt{2} (\Delta +\varepsilon ) x_{12}+t_r x_{13}-t_l x_{22}\right)}{2\hbar} &
		\frac{\I \sqrt{3} m \left(t_r x_{22}-t_l x_{31}+\sqrt{2} (\Delta -\varepsilon ) x_{32}-t_r x_{33}\right)}{2\hbar} \\
	\end{array}\right),\\
	\bo{C} & \equiv \left(\begin{array}{cc}
		-\frac{\I \sqrt{2} m t_l (x_{12}-x_{21})}{\hbar} &
		-\frac{\I m \left(\sqrt{2} t_r x_{21}-4 \varepsilon  x_{31}-\sqrt{2} t_l x_{32}\right)}{2\hbar} \\
		\frac{\I m \left(\sqrt{2} t_r x_{12}-4 \varepsilon  x_{13}-\sqrt{2} t_l x_{23}\right)}{2\hbar} &
		\frac{\I \sqrt{2} m t_r (x_{23}-x_{32})}{\hbar} \\
	\end{array}\right).
	\end{split}
	\label{eq:app_momentum_full}
\end{align}

In the case of real-valued matrix elements $ x_{ij}=x_{ji}\in\mathbb{R} $ with $ i,j\in\lbrace 1,2,3\rbrace $ in Eq.~\eqref{eq:app_momentum_full} we obtain a simpler expression for the parameters in Eq.~\eqref{eq:app_momentum_help},
\begin{align}
\begin{split}
	\bo{A_R} &\equiv \left(\begin{array}{cc}
		0 & 0\\
		0 & 0\\
	\end{array}\right),\\
	\bo{B_R} & \equiv -\left(\begin{array}{cc}
		\frac{\I m \left(t_l x_{11}-\sqrt{2} (\Delta +\varepsilon ) x_{12}-t_r x_{13}-t_l x_{22}\right)}{2\hbar} &
		-\frac{\I m \left(t_r x_{22}+t_l x_{31}+\sqrt{2} (\Delta -\varepsilon ) x_{32}-t_r x_{33}\right)}{2\hbar} \\
		\frac{\I \sqrt{3} m \left(t_l x_{11}-\sqrt{2} (\Delta +\varepsilon ) x_{12}+t_r x_{13}-t_l x_{22}\right)}{2\hbar} &
		\frac{\I \sqrt{3} m \left(t_r x_{22}-t_l x_{31}+\sqrt{2} (\Delta -\varepsilon ) x_{32}-t_r x_{33}\right)}{2\hbar} \\
	\end{array}\right),\\
	\bo{C_R} & \equiv \left(\begin{array}{cc}
		0 &
		-\frac{\I m \left(\sqrt{2} t_r x_{21}-4 \varepsilon  x_{31}-\sqrt{2} t_l x_{32}\right)}{2\hbar} \\
		\frac{\I m \left(\sqrt{2} t_r x_{12}-4 \varepsilon  x_{13}-\sqrt{2} t_l x_{23}\right)}{2\hbar} &
		0 \\
	\end{array}\right).
	\end{split}
	\label{eq:app_momentum_simp}
\end{align}

Shifting the momentum operator into the qubit basis of the RX regime acquired through a second-order Schrieffer-Wolff transformation\cite{Taylor2013},  $ \tilde{\bo{p}} = \E^{S}\bo{p}\E^{-S} \approx \bo{p} - \left[\bo{p},S\right] $ with the anti-hermitian matrix $S$. We obtain 
\begin{align}
\begin{split}
	\tilde{\bo{p}}\approx\bo{p} - \left[\bo{p},S\right] =\left(\begin{array}{cc}
		A +sB^{\dagger}+Bs^{\dagger} & +B -As+sC\\
		B^{\dagger}-s^{\dagger}A +C s^{\dagger}C & C - s^{\dagger} B - B^{\dagger} s \\
	\end{array}\right),
\end{split}
\end{align}
where $s$ is the $2\times 2$ block in the top-right corner of matrix $S$. Inserting the expressions for $A$, $B$, $C$, and $s$ we obtain the expression $\tilde{\bo{p}}|_{\ket{0},\ket{1}}\approx A +sB^{\dagger}+Bs^{\dagger}$ given in Eq.~\eqref{eq:matrix_elements}.

\section{Calculation of the coupling strength $ g_r $}
\label{app:calc_coup}
In order to obtain the qubit-cavity coupling $g_{r}$, we have to calculate the matrix elements $x_{ij}$ in Eq.~\eqref{eq:matrix_elements} with $i,j\in \lbrace 1,2,3\rbrace$ which we can rewrite due to a transformation back into the non-orthogonal basis of $\lbrace\ket{\varphi_{1}},\ket{\varphi_{2}},\ket{\varphi_{3}}\rbrace$ as $x_{ij}=\sum_{k,l=1}^{3} q_{kl}\bra{\varphi_{k}} x \ket{\varphi_{l}}$ with $k,l\in\lbrace 1,2,3 \rbrace$ and $q_{kl}\in \mathbb{R}$. Due to the separability of the Fock-Darwin wave functions in position space, $\varphi_{k}(x,y)=\varphi_{k,x}(x)\varphi_{k,y}(y)$, where $\varphi_{k,x}(x)\in\mathbb{R}$, the integrations over both space dimensions are independent,
\begin{align}
\bra{\varphi_{k}} \hat{x}  \ket{\varphi_{l}} =\iint\limits_{-\infty}^{\quad\infty}dx\,dy\,\varphi_{k}^{*}(x,y)\,x\,\varphi_{l}(x,y) = \left(\int\limits_{-\infty}^{\infty} dx \varphi_{k,x}(x)\,x\,\varphi_{l,x}(x)\right)\left(\int\limits_{-\infty}^{\infty} dy \varphi_{k,y}^{*}(y)\varphi_{l,y}(y)\right).
\end{align}
In the next step we prove that this integral is always real under the assumption that the confinement energy $\omega\st{QD}$ is identical in each QD, even for a finite homogenous magnetic field. For the integration over $x$, this is true, since the wave functions do not acquire a phase in $x$-direction, hence, it has no imaginary contribution. Proving this statement for the second term which is just the overlap in $y$-direction, is more complicated due to the additional phase acquired by the magnetic field. We obtain, e.g., for $k=1$ and $l=2$ for the second term,
\begin{align}
\begin{split}
\int\limits_{-\infty}^{\infty} dy\, \varphi_{k,y}^{*}(y)\varphi_{l,y}(y) &= \int\limits_{-\infty}^{\infty} dy\, \E^{\I a_{l}y/2l_{B}^{2}}\E^{-y^{2}/8a_{S}} 
 = \int\limits_{-\infty}^{\infty} d\tilde{y}\,\E^{- \tilde{y}^{2}/8a_{S}^{2}}\, \E^{- a_{l}^{2}a_{S}^{2}/2l_{B}^{4}},
\end{split}
\end{align}
with $\tilde{y}=y-2\im a_{l}a_{S}^{2}/l_{B}^{2}$, which is a Gaussian integral shifted along the imaginary axis. Since this integral is real, $x_{ij}$ is also real and the only relevant matrix element according to Eq.~\eqref{eq:matrix_elements} is
\begin{align}
	 x_{13}  = \frac{1}{N_1 \, N_3} &\left(\bra{\varphi_1} + a_1\bra{\varphi_2} + b_1\bra{\varphi_3}\right) \hat{x} \left(\ket{\varphi_3} + a_3\ket{\varphi_2} + b_3\ket{\varphi_1} \right).
\end{align}
To calculate $x_{13}$, we first consider the more general case of non-vanishing overlap $ S_{13} $ for our calculation and neglect $S_{13}$ later. Inserting the Fock-Darwin functions from Eq.~\eqref{eq:FockDarwin}, we find
\begin{align}
\begin{split}
x_{13}  = \frac{m\omega\st{QD}}{\pi  \hbar N_1 N_3}\iint\limits_{-\infty}^{\quad\infty}& dx\,dy\,  \left(a_1 \exp\left[-\frac{m\omega\st{QD} \left(x^2+y^2\right)}{2\hbar}\right]+\exp \left\lbrace-\frac{m\omega\st{QD} \left[(a_l+x)^2+y^2\right]}{2\hbar}+\frac{i a_l y}{2 l_B^2}\right\rbrace\right.\\ 
 &\left.\quad + b_1 \exp \left\lbrace-\frac{m\omega\st{QD} \left[(a_r-x)^2+y^2\right]}{2\hbar}-\frac{i a_r y}{2 l_B^2}\right\rbrace\right)x\\ 
 &\times\left(a_3\exp\left[-\frac{m\omega\st{QD} \left(x^2+y^2\right)}{2\hbar}\right]+b_3\exp \left\lbrace-\frac{m\omega\st{QD} \left[(a_l+x)^2+y^2\right]}{2\hbar}-\frac{i a_l y}{2 l_B^2}\right\rbrace\right.\\
 &\left.\quad +\exp \left\lbrace-\frac{m\omega\st{QD} \left[(a_r-x)^2+y^2\right]}{2\hbar}+\frac{i a_r y}{2 l_B^2}\right\rbrace\right).
 \end{split}
\end{align}
\end{widetext}
This integral can be solved straight-forwardly, since all integrands are Gaussian. As a result, we obtain
\begin{align}
\begin{split}
x_{13}  =& \left[ - a_l (a_1 b_3 S_l+a_3S_l+2 b_3+S_{13})+a_r (a_1S_r+S_{13})\right. \\
&+\left. b_1a_r (a_3 S_r+2 b_3+2)\right] /2 N_1 N_3.
\end{split}
\end{align}
Setting $ S_{13}=b_1=b_3=0 $ and inserting $a_{1}$, $a_{2}$, $b_{2}$ and $a_{3}$ from Section~\ref{ssec:OWOa} for the second step, we obtain
\begin{align}
\begin{split}
x_{13}  &= \frac{a_1 a_r S_r-a_3 a_l S_l}{2 \sqrt{a_1^2+2 a_1 S_l+1} \sqrt{a_3^2+2 a_3S_r+1}}\\
						&= \tilde{a}_\text{rel} (a_l-a_{r})S_{l} \, S_{r},
\end{split}
\end{align}
where $\tilde{a}_\text{rel}$ can be expanded for $k =a_{1}S_{r}/a_{3}S_{l}\approx 1$ (defined and discussed in Appendix~\ref{app:mini}),
\begin{align}
\tilde{a}_\text{rel}\simeq 1+\left(\frac{a_{l}}{a_{r} - a_{l}}-S_l^2+S_r^2+\frac{1}{2}\right)(k-1)
\end{align}
which yields the expression in Eq.~\eqref{eq:mom_ULD} after inserting into Eq.~\eqref{eq:matrix_elements}.

\section{Second SW transformation to eliminate the cavity mode.}
\label{app:SW2}

In this appendix, we present an effective Hamiltonian in which the cavity mode is split off by a second SW transformation in order to express the cavity mediated coupling between two RX qubits.\cite{Burkard2006} We start with the universal Hamiltonian in the rotating frame from Eq.~\eqref{eq:rotatingwave} and introduce a second order SW transformation 
\begin{align}
	H\st{tot}\equiv\E^{S_2} H\st{RW} E^{-S_2} \simeq H_0 + \left[H\st{int},S_2\right]/2.
\end{align}
where $ H_0 \equiv \sum_i \hbar\omega_i/2\,\sigma_z +\hbar\omega\st{ph}a^\dagger a  $ and $ H\st{int} = \sum_i g_i\,\I \left(\sigma_+a-\sigma_-a^\dagger\right) $ and the anti-Hermitian operator $ S_2 $ is determined by the SW condition $ \left[H_0,S_2\right] = -H\st{int} $. Hence, we express $ S_2 $ in terms of operators,
\begin{align}
	S_2 = \sum\limits_{i=1}^{2} \frac{g_i}{\hbar\omega\st{ph}-\hbar\omega_i}\left(\sigma_+a-\sigma_-a^\dagger\right),
\end{align}
with the previously introduced ladder operators $ \sigma_\pm $.

\bibliography{lit2}

\begin{thebibliography}{53}%
\makeatletter
\providecommand \@ifxundefined [1]{%
 \@ifx{#1\undefined}
}%
\providecommand \@ifnum [1]{%
 \ifnum #1\expandafter \@firstoftwo
 \else \expandafter \@secondoftwo
 \fi
}%
\providecommand \@ifx [1]{%
 \ifx #1\expandafter \@firstoftwo
 \else \expandafter \@secondoftwo
 \fi
}%
\providecommand \natexlab [1]{#1}%
\providecommand \enquote  [1]{``#1''}%
\providecommand \bibnamefont  [1]{#1}%
\providecommand \bibfnamefont [1]{#1}%
\providecommand \citenamefont [1]{#1}%
\providecommand \href@noop [0]{\@secondoftwo}%
\providecommand \href [0]{\begingroup \@sanitize@url \@href}%
\providecommand \@href[1]{\@@startlink{#1}\@@href}%
\providecommand \@@href[1]{\endgroup#1\@@endlink}%
\providecommand \@sanitize@url [0]{\catcode `\\12\catcode `\$12\catcode
  `\&12\catcode `\#12\catcode `\^12\catcode `\_12\catcode `\%12\relax}%
\providecommand \@@startlink[1]{}%
\providecommand \@@endlink[0]{}%
\providecommand \url  [0]{\begingroup\@sanitize@url \@url }%
\providecommand \@url [1]{\endgroup\@href {#1}{\urlprefix }}%
\providecommand \urlprefix  [0]{URL }%
\providecommand \Eprint [0]{\href }%
\providecommand \doibase [0]{http://dx.doi.org/}%
\providecommand \selectlanguage [0]{\@gobble}%
\providecommand \bibinfo  [0]{\@secondoftwo}%
\providecommand \bibfield  [0]{\@secondoftwo}%
\providecommand \translation [1]{[#1]}%
\providecommand \BibitemOpen [0]{}%
\providecommand \bibitemStop [0]{}%
\providecommand \bibitemNoStop [0]{.\EOS\space}%
\providecommand \EOS [0]{\spacefactor3000\relax}%
\providecommand \BibitemShut  [1]{\csname bibitem#1\endcsname}%
\let\auto@bib@innerbib\@empty
%</preamble>
\bibitem [{\citenamefont {Loss}\ and\ \citenamefont
  {DiVincenzo}(1998)}]{PhysRevA.57.120}%
  \BibitemOpen
  \bibfield  {author} {\bibinfo {author} {\bibfnamefont {D.}~\bibnamefont
  {Loss}}\ and\ \bibinfo {author} {\bibfnamefont {D.~P.}\ \bibnamefont
  {DiVincenzo}},\ }\href {\doibase 10.1103/PhysRevA.57.120} {\bibfield
  {journal} {\bibinfo  {journal} {Phys. Rev. A}\ }\textbf {\bibinfo {volume}
  {57}},\ \bibinfo {pages} {120} (\bibinfo {year} {1998})}\BibitemShut
  {NoStop}%
\bibitem [{\citenamefont {Petta}\ \emph {et~al.}(2005)\citenamefont {Petta},
  \citenamefont {Johnson}, \citenamefont {Taylor}, \citenamefont {Laird},
  \citenamefont {Yacoby}, \citenamefont {Lukin}, \citenamefont {Marcus},
  \citenamefont {Hanson},\ and\ \citenamefont {Gossard}}]{Petta2005}%
  \BibitemOpen
  \bibfield  {author} {\bibinfo {author} {\bibfnamefont {J.~R.}\ \bibnamefont
  {Petta}}, \bibinfo {author} {\bibfnamefont {A.~C.}\ \bibnamefont {Johnson}},
  \bibinfo {author} {\bibfnamefont {J.~M.}\ \bibnamefont {Taylor}}, \bibinfo
  {author} {\bibfnamefont {E.~A.}\ \bibnamefont {Laird}}, \bibinfo {author}
  {\bibfnamefont {A.}~\bibnamefont {Yacoby}}, \bibinfo {author} {\bibfnamefont
  {M.~D.}\ \bibnamefont {Lukin}}, \bibinfo {author} {\bibfnamefont {C.~M.}\
  \bibnamefont {Marcus}}, \bibinfo {author} {\bibfnamefont {M.~P.}\
  \bibnamefont {Hanson}}, \ and\ \bibinfo {author} {\bibfnamefont {A.~C.}\
  \bibnamefont {Gossard}},\ }\href {\doibase 10.1126/science.1116955}
  {\bibfield  {journal} {\bibinfo  {journal} {Science}\ }\textbf {\bibinfo
  {volume} {309}},\ \bibinfo {pages} {2180} (\bibinfo {year}
  {2005})}\BibitemShut {NoStop}%
\bibitem [{\citenamefont {Greilich}\ \emph {et~al.}(2006)\citenamefont
  {Greilich}, \citenamefont {Yakovlev}, \citenamefont {Shabaev}, \citenamefont
  {Efros}, \citenamefont {Yugova}, \citenamefont {Oulton}, \citenamefont
  {Stavarache}, \citenamefont {Reuter}, \citenamefont {Wieck},\ and\
  \citenamefont {Bayer}}]{Greilich2006}%
  \BibitemOpen
  \bibfield  {author} {\bibinfo {author} {\bibfnamefont {A.}~\bibnamefont
  {Greilich}}, \bibinfo {author} {\bibfnamefont {D.~R.}\ \bibnamefont
  {Yakovlev}}, \bibinfo {author} {\bibfnamefont {A.}~\bibnamefont {Shabaev}},
  \bibinfo {author} {\bibfnamefont {A.~L.}\ \bibnamefont {Efros}}, \bibinfo
  {author} {\bibfnamefont {I.~A.}\ \bibnamefont {Yugova}}, \bibinfo {author}
  {\bibfnamefont {R.}~\bibnamefont {Oulton}}, \bibinfo {author} {\bibfnamefont
  {V.}~\bibnamefont {Stavarache}}, \bibinfo {author} {\bibfnamefont
  {D.}~\bibnamefont {Reuter}}, \bibinfo {author} {\bibfnamefont
  {A.}~\bibnamefont {Wieck}}, \ and\ \bibinfo {author} {\bibfnamefont
  {M.}~\bibnamefont {Bayer}},\ }\href {\doibase 10.1126/science.1128215}
  {\bibfield  {journal} {\bibinfo  {journal} {Science}\ }\textbf {\bibinfo
  {volume} {313}},\ \bibinfo {pages} {341} (\bibinfo {year}
  {2006})}\BibitemShut {NoStop}%
\bibitem [{\citenamefont {Koppens}\ \emph {et~al.}(2008)\citenamefont
  {Koppens}, \citenamefont {Nowack},\ and\ \citenamefont
  {Vandersypen}}]{Koppens2008}%
  \BibitemOpen
  \bibfield  {author} {\bibinfo {author} {\bibfnamefont {F.~H.~L.}\
  \bibnamefont {Koppens}}, \bibinfo {author} {\bibfnamefont {K.~C.}\
  \bibnamefont {Nowack}}, \ and\ \bibinfo {author} {\bibfnamefont {L.~M.~K.}\
  \bibnamefont {Vandersypen}},\ }\href {\doibase
  10.1103/PhysRevLett.100.236802} {\bibfield  {journal} {\bibinfo  {journal}
  {Phys. Rev. Lett.}\ }\textbf {\bibinfo {volume} {100}},\ \bibinfo {pages}
  {236802} (\bibinfo {year} {2008})}\BibitemShut {NoStop}%
\bibitem [{\citenamefont {Bluhm}\ \emph {et~al.}(2011)\citenamefont {Bluhm},
  \citenamefont {Foletti}, \citenamefont {Neder}, \citenamefont {Rudner},
  \citenamefont {Mahalu}, \citenamefont {Umansky},\ and\ \citenamefont
  {Yacoby}}]{Bluhm2011}%
  \BibitemOpen
  \bibfield  {author} {\bibinfo {author} {\bibfnamefont {H.}~\bibnamefont
  {Bluhm}}, \bibinfo {author} {\bibfnamefont {S.}~\bibnamefont {Foletti}},
  \bibinfo {author} {\bibfnamefont {I.}~\bibnamefont {Neder}}, \bibinfo
  {author} {\bibfnamefont {M.}~\bibnamefont {Rudner}}, \bibinfo {author}
  {\bibfnamefont {D.}~\bibnamefont {Mahalu}}, \bibinfo {author} {\bibfnamefont
  {V.}~\bibnamefont {Umansky}}, \ and\ \bibinfo {author} {\bibfnamefont
  {A.}~\bibnamefont {Yacoby}},\ }\href {\doibase 10.1038/nphys1856} {\bibfield
  {journal} {\bibinfo  {journal} {Nat Phys}\ }\textbf {\bibinfo {volume} {7}},\
  \bibinfo {pages} {109} (\bibinfo {year} {2011})}\BibitemShut {NoStop}%
\bibitem [{\citenamefont {Hanson}\ \emph {et~al.}(2007)\citenamefont {Hanson},
  \citenamefont {Kouwenhoven}, \citenamefont {Petta}, \citenamefont {Tarucha},\
  and\ \citenamefont {Vandersypen}}]{Hanson2007}%
  \BibitemOpen
  \bibfield  {author} {\bibinfo {author} {\bibfnamefont {R.}~\bibnamefont
  {Hanson}}, \bibinfo {author} {\bibfnamefont {L.~P.}\ \bibnamefont
  {Kouwenhoven}}, \bibinfo {author} {\bibfnamefont {J.~R.}\ \bibnamefont
  {Petta}}, \bibinfo {author} {\bibfnamefont {S.}~\bibnamefont {Tarucha}}, \
  and\ \bibinfo {author} {\bibfnamefont {L.~M.~K.}\ \bibnamefont
  {Vandersypen}},\ }\href {\doibase 10.1103/RevModPhys.79.1217} {\bibfield
  {journal} {\bibinfo  {journal} {Rev. Mod. Phys.}\ }\textbf {\bibinfo {volume}
  {79}},\ \bibinfo {pages} {1217} (\bibinfo {year} {2007})}\BibitemShut
  {NoStop}%
\bibitem [{\citenamefont {Awschalom}\ \emph {et~al.}(2013)\citenamefont
  {Awschalom}, \citenamefont {Bassett}, \citenamefont {Dzurak}, \citenamefont
  {Hu},\ and\ \citenamefont {Petta}}]{Awschalom2013}%
  \BibitemOpen
  \bibfield  {author} {\bibinfo {author} {\bibfnamefont {D.~D.}\ \bibnamefont
  {Awschalom}}, \bibinfo {author} {\bibfnamefont {L.~C.}\ \bibnamefont
  {Bassett}}, \bibinfo {author} {\bibfnamefont {A.~S.}\ \bibnamefont {Dzurak}},
  \bibinfo {author} {\bibfnamefont {E.~L.}\ \bibnamefont {Hu}}, \ and\ \bibinfo
  {author} {\bibfnamefont {J.~R.}\ \bibnamefont {Petta}},\ }\href {\doibase
  10.1126/science.1231364} {\bibfield  {journal} {\bibinfo  {journal}
  {Science}\ }\textbf {\bibinfo {volume} {339}},\ \bibinfo {pages} {1174}
  (\bibinfo {year} {2013})}\BibitemShut {NoStop}%
\bibitem [{\citenamefont {Zwanenburg}\ \emph {et~al.}(2013)\citenamefont
  {Zwanenburg}, \citenamefont {Dzurak}, \citenamefont {Morello}, \citenamefont
  {Simmons}, \citenamefont {Hollenberg}, \citenamefont {Klimeck}, \citenamefont
  {Rogge}, \citenamefont {Coppersmith},\ and\ \citenamefont
  {Eriksson}}]{Zwanenburg2013}%
  \BibitemOpen
  \bibfield  {author} {\bibinfo {author} {\bibfnamefont {F.~A.}\ \bibnamefont
  {Zwanenburg}}, \bibinfo {author} {\bibfnamefont {A.~S.}\ \bibnamefont
  {Dzurak}}, \bibinfo {author} {\bibfnamefont {A.}~\bibnamefont {Morello}},
  \bibinfo {author} {\bibfnamefont {M.~Y.}\ \bibnamefont {Simmons}}, \bibinfo
  {author} {\bibfnamefont {L.~C.~L.}\ \bibnamefont {Hollenberg}}, \bibinfo
  {author} {\bibfnamefont {G.}~\bibnamefont {Klimeck}}, \bibinfo {author}
  {\bibfnamefont {S.}~\bibnamefont {Rogge}}, \bibinfo {author} {\bibfnamefont
  {S.~N.}\ \bibnamefont {Coppersmith}}, \ and\ \bibinfo {author} {\bibfnamefont
  {M.~A.}\ \bibnamefont {Eriksson}},\ }\href {\doibase
  10.1103/RevModPhys.85.961} {\bibfield  {journal} {\bibinfo  {journal} {Rev.
  Mod. Phys.}\ }\textbf {\bibinfo {volume} {85}},\ \bibinfo {pages} {961}
  (\bibinfo {year} {2013})}\BibitemShut {NoStop}%
\bibitem [{\citenamefont {Petersson}\ \emph {et~al.}(2010)\citenamefont
  {Petersson}, \citenamefont {Petta}, \citenamefont {Lu},\ and\ \citenamefont
  {Gossard}}]{Petersson2010}%
  \BibitemOpen
  \bibfield  {author} {\bibinfo {author} {\bibfnamefont {K.}~\bibnamefont
  {Petersson}}, \bibinfo {author} {\bibfnamefont {J.}~\bibnamefont {Petta}},
  \bibinfo {author} {\bibfnamefont {H.}~\bibnamefont {Lu}}, \ and\ \bibinfo
  {author} {\bibfnamefont {A.}~\bibnamefont {Gossard}},\ }\href {\doibase
  10.1103/PhysRevLett.105.246804} {\bibfield  {journal} {\bibinfo  {journal}
  {Phys. Rev. Lett.}\ }\textbf {\bibinfo {volume} {105}},\ \bibinfo {pages}
  {246804} (\bibinfo {year} {2010})}\BibitemShut {NoStop}%
\bibitem [{\citenamefont {Shi}\ \emph {et~al.}(2013)\citenamefont {Shi},
  \citenamefont {Simmons}, \citenamefont {Ward}, \citenamefont {Prance},
  \citenamefont {Mohr}, \citenamefont {Koh}, \citenamefont {Gamble},
  \citenamefont {Wu}, \citenamefont {Savage}, \citenamefont {Lagally},
  \citenamefont {Friesen}, \citenamefont {Coppersmith},\ and\ \citenamefont
  {Eriksson}}]{Shi2013}%
  \BibitemOpen
  \bibfield  {author} {\bibinfo {author} {\bibfnamefont {Z.}~\bibnamefont
  {Shi}}, \bibinfo {author} {\bibfnamefont {C.~B.}\ \bibnamefont {Simmons}},
  \bibinfo {author} {\bibfnamefont {D.~R.}\ \bibnamefont {Ward}}, \bibinfo
  {author} {\bibfnamefont {J.~R.}\ \bibnamefont {Prance}}, \bibinfo {author}
  {\bibfnamefont {R.~T.}\ \bibnamefont {Mohr}}, \bibinfo {author}
  {\bibfnamefont {T.~S.}\ \bibnamefont {Koh}}, \bibinfo {author} {\bibfnamefont
  {J.~K.}\ \bibnamefont {Gamble}}, \bibinfo {author} {\bibfnamefont
  {X.}~\bibnamefont {Wu}}, \bibinfo {author} {\bibfnamefont {D.~E.}\
  \bibnamefont {Savage}}, \bibinfo {author} {\bibfnamefont {M.~G.}\
  \bibnamefont {Lagally}}, \bibinfo {author} {\bibfnamefont {M.}~\bibnamefont
  {Friesen}}, \bibinfo {author} {\bibfnamefont {S.~N.}\ \bibnamefont
  {Coppersmith}}, \ and\ \bibinfo {author} {\bibfnamefont {M.~A.}\ \bibnamefont
  {Eriksson}},\ }\href {\doibase 10.1103/PhysRevB.88.075416} {\bibfield
  {journal} {\bibinfo  {journal} {Phys. Rev. B}\ }\textbf {\bibinfo {volume}
  {88}},\ \bibinfo {pages} {075416} (\bibinfo {year} {2013})}\BibitemShut
  {NoStop}%
\bibitem [{\citenamefont {Kim}\ \emph {et~al.}(2015)\citenamefont {Kim},
  \citenamefont {{D. R. Ward}}, \citenamefont {{C. B. Simmons}}, \citenamefont
  {Gamble}, \citenamefont {Blume-Kohout}, \citenamefont {Nielsen},
  \citenamefont {{D. E. Savage}}, \citenamefont {{M. G. Lagally}},
  \citenamefont {Friesen}, \citenamefont {{S. N. Coppersmith}},\ and\
  \citenamefont {{M. A. Eriksson}}}]{Kim2015}%
  \BibitemOpen
  \bibfield  {author} {\bibinfo {author} {\bibfnamefont {D.}~\bibnamefont
  {Kim}}, \bibinfo {author} {\bibnamefont {{D. R. Ward}}}, \bibinfo {author}
  {\bibnamefont {{C. B. Simmons}}}, \bibinfo {author} {\bibfnamefont {J.~K.}\
  \bibnamefont {Gamble}}, \bibinfo {author} {\bibfnamefont {R.}~\bibnamefont
  {Blume-Kohout}}, \bibinfo {author} {\bibfnamefont {E.}~\bibnamefont
  {Nielsen}}, \bibinfo {author} {\bibnamefont {{D. E. Savage}}}, \bibinfo
  {author} {\bibnamefont {{M. G. Lagally}}}, \bibinfo {author} {\bibfnamefont
  {M.}~\bibnamefont {Friesen}}, \bibinfo {author} {\bibnamefont {{S. N.
  Coppersmith}}}, \ and\ \bibinfo {author} {\bibnamefont {{M. A. Eriksson}}},\
  }\href {http://dx.doi.org/10.1038/nnano.2014.336} {\bibfield  {journal}
  {\bibinfo  {journal} {Nat Nano}\ }\textbf {\bibinfo {volume} {10}},\ \bibinfo
  {pages} {243} (\bibinfo {year} {2015})}\BibitemShut {NoStop}%
\bibitem [{\citenamefont {Yurkevich}\ \emph {et~al.}(2010)\citenamefont
  {Yurkevich}, \citenamefont {Baldwin}, \citenamefont {Lerner},\ and\
  \citenamefont {Altshuler}}]{Yurkevich2010}%
  \BibitemOpen
  \bibfield  {author} {\bibinfo {author} {\bibfnamefont {I.~V.}\ \bibnamefont
  {Yurkevich}}, \bibinfo {author} {\bibfnamefont {J.}~\bibnamefont {Baldwin}},
  \bibinfo {author} {\bibfnamefont {I.~V.}\ \bibnamefont {Lerner}}, \ and\
  \bibinfo {author} {\bibfnamefont {B.~L.}\ \bibnamefont {Altshuler}},\ }\href
  {\doibase 10.1103/PhysRevB.81.121305} {\bibfield  {journal} {\bibinfo
  {journal} {Phys. Rev. B}\ }\textbf {\bibinfo {volume} {81}},\ \bibinfo
  {pages} {121305} (\bibinfo {year} {2010})}\BibitemShut {NoStop}%
\bibitem [{\citenamefont {Gaudreau}\ \emph {et~al.}(2006)\citenamefont
  {Gaudreau}, \citenamefont {Studenikin}, \citenamefont {Sachrajda},
  \citenamefont {Zawadzki}, \citenamefont {Kam}, \citenamefont {Lapointe},
  \citenamefont {Korkusinski},\ and\ \citenamefont {Hawrylak}}]{Gaudreau2006}%
  \BibitemOpen
  \bibfield  {author} {\bibinfo {author} {\bibfnamefont {L.}~\bibnamefont
  {Gaudreau}}, \bibinfo {author} {\bibfnamefont {S.~A.}\ \bibnamefont
  {Studenikin}}, \bibinfo {author} {\bibfnamefont {A.~S.}\ \bibnamefont
  {Sachrajda}}, \bibinfo {author} {\bibfnamefont {P.}~\bibnamefont {Zawadzki}},
  \bibinfo {author} {\bibfnamefont {A.}~\bibnamefont {Kam}}, \bibinfo {author}
  {\bibfnamefont {J.}~\bibnamefont {Lapointe}}, \bibinfo {author}
  {\bibfnamefont {M.}~\bibnamefont {Korkusinski}}, \ and\ \bibinfo {author}
  {\bibfnamefont {P.}~\bibnamefont {Hawrylak}},\ }\href {\doibase
  10.1103/PhysRevLett.97.036807} {\bibfield  {journal} {\bibinfo  {journal}
  {Phys. Rev. Lett.}\ }\textbf {\bibinfo {volume} {97}},\ \bibinfo {pages}
  {036807} (\bibinfo {year} {2006})}\BibitemShut {NoStop}%
\bibitem [{\citenamefont {Takakura}\ \emph {et~al.}(2010)\citenamefont
  {Takakura}, \citenamefont {Pioro-Ladri\`{e}re}, \citenamefont {Obata},
  \citenamefont {Shin}, \citenamefont {Brunner}, \citenamefont {Yoshida},
  \citenamefont {Taniyama},\ and\ \citenamefont {Tarucha}}]{Takakura2010}%
  \BibitemOpen
  \bibfield  {author} {\bibinfo {author} {\bibfnamefont {T.}~\bibnamefont
  {Takakura}}, \bibinfo {author} {\bibfnamefont {M.}~\bibnamefont
  {Pioro-Ladri\`{e}re}}, \bibinfo {author} {\bibfnamefont {T.}~\bibnamefont
  {Obata}}, \bibinfo {author} {\bibfnamefont {Y.-S.}\ \bibnamefont {Shin}},
  \bibinfo {author} {\bibfnamefont {R.}~\bibnamefont {Brunner}}, \bibinfo
  {author} {\bibfnamefont {K.}~\bibnamefont {Yoshida}}, \bibinfo {author}
  {\bibfnamefont {T.}~\bibnamefont {Taniyama}}, \ and\ \bibinfo {author}
  {\bibfnamefont {S.}~\bibnamefont {Tarucha}},\ }\href {\doibase
  http://dx.doi.org/10.1063/1.3518919} {\bibfield  {journal} {\bibinfo
  {journal} {Appl. Phys. Lett.}\ }\textbf {\bibinfo {volume} {97}},\ \bibinfo
  {eid} {212104} (\bibinfo {year} {2010})}\BibitemShut {NoStop}%
\bibitem [{\citenamefont {Gaudreau}\ \emph {et~al.}(2012)\citenamefont
  {Gaudreau}, \citenamefont {Granger}, \citenamefont {Kam}, \citenamefont
  {Aers}, \citenamefont {Studenikin}, \citenamefont {Zawadzki}, \citenamefont
  {Pioro-Ladriere}, \citenamefont {Wasilewski},\ and\ \citenamefont
  {Sachrajda}}]{Gaudreau2012}%
  \BibitemOpen
  \bibfield  {author} {\bibinfo {author} {\bibfnamefont {L.}~\bibnamefont
  {Gaudreau}}, \bibinfo {author} {\bibfnamefont {G.}~\bibnamefont {Granger}},
  \bibinfo {author} {\bibfnamefont {A.}~\bibnamefont {Kam}}, \bibinfo {author}
  {\bibfnamefont {G.~C.}\ \bibnamefont {Aers}}, \bibinfo {author}
  {\bibfnamefont {S.~A.}\ \bibnamefont {Studenikin}}, \bibinfo {author}
  {\bibfnamefont {P.}~\bibnamefont {Zawadzki}}, \bibinfo {author}
  {\bibfnamefont {M.}~\bibnamefont {Pioro-Ladriere}}, \bibinfo {author}
  {\bibfnamefont {Z.~R.}\ \bibnamefont {Wasilewski}}, \ and\ \bibinfo {author}
  {\bibfnamefont {A.~S.}\ \bibnamefont {Sachrajda}},\ }\href
  {http://dx.doi.org/10.1038/nphys2149} {\bibfield  {journal} {\bibinfo
  {journal} {Nat Phys}\ }\textbf {\bibinfo {volume} {8}},\ \bibinfo {pages}
  {54} (\bibinfo {year} {2012})}\BibitemShut {NoStop}%
\bibitem [{\citenamefont {Medford}\ \emph
  {et~al.}(2013{\natexlab{a}})\citenamefont {Medford}, \citenamefont {Beil},
  \citenamefont {Taylor}, \citenamefont {Bartlett}, \citenamefont {Doherty},
  \citenamefont {Rashba}, \citenamefont {DiVincenzo}, \citenamefont {Lu},
  \citenamefont {Gossard},\ and\ \citenamefont {Marcus}}]{Medford2013N}%
  \BibitemOpen
  \bibfield  {author} {\bibinfo {author} {\bibfnamefont {J.}~\bibnamefont
  {Medford}}, \bibinfo {author} {\bibfnamefont {J.}~\bibnamefont {Beil}},
  \bibinfo {author} {\bibfnamefont {J.~M.}\ \bibnamefont {Taylor}}, \bibinfo
  {author} {\bibfnamefont {S.~D.}\ \bibnamefont {Bartlett}}, \bibinfo {author}
  {\bibfnamefont {A.~C.}\ \bibnamefont {Doherty}}, \bibinfo {author}
  {\bibfnamefont {E.~I.}\ \bibnamefont {Rashba}}, \bibinfo {author}
  {\bibfnamefont {D.~P.}\ \bibnamefont {DiVincenzo}}, \bibinfo {author}
  {\bibfnamefont {H.}~\bibnamefont {Lu}}, \bibinfo {author} {\bibfnamefont
  {A.~C.}\ \bibnamefont {Gossard}}, \ and\ \bibinfo {author} {\bibfnamefont
  {C.~M.}\ \bibnamefont {Marcus}},\ }\href
  {http://dx.doi.org/10.1038/nnano.2013.168} {\bibfield  {journal} {\bibinfo
  {journal} {Nat Nano}\ }\textbf {\bibinfo {volume} {8}},\ \bibinfo {pages}
  {654} (\bibinfo {year} {2013}{\natexlab{a}})}\BibitemShut {NoStop}%
\bibitem [{\citenamefont {Medford}\ \emph
  {et~al.}(2013{\natexlab{b}})\citenamefont {Medford}, \citenamefont {Beil},
  \citenamefont {Taylor}, \citenamefont {Rashba}, \citenamefont {Lu},
  \citenamefont {Gossard},\ and\ \citenamefont {Marcus}}]{Medford2013}%
  \BibitemOpen
  \bibfield  {author} {\bibinfo {author} {\bibfnamefont {J.}~\bibnamefont
  {Medford}}, \bibinfo {author} {\bibfnamefont {J.}~\bibnamefont {Beil}},
  \bibinfo {author} {\bibfnamefont {J.~M.}\ \bibnamefont {Taylor}}, \bibinfo
  {author} {\bibfnamefont {E.~I.}\ \bibnamefont {Rashba}}, \bibinfo {author}
  {\bibfnamefont {H.}~\bibnamefont {Lu}}, \bibinfo {author} {\bibfnamefont
  {A.~C.}\ \bibnamefont {Gossard}}, \ and\ \bibinfo {author} {\bibfnamefont
  {C.~M.}\ \bibnamefont {Marcus}},\ }\href {\doibase
  10.1103/PhysRevLett.111.050501} {\bibfield  {journal} {\bibinfo  {journal}
  {Phys. Rev. Lett.}\ }\textbf {\bibinfo {volume} {111}},\ \bibinfo {pages}
  {050501} (\bibinfo {year} {2013}{\natexlab{b}})}\BibitemShut {NoStop}%
\bibitem [{\citenamefont {Veldhorst}\ \emph {et~al.}(2015)\citenamefont
  {Veldhorst}, \citenamefont {Yang}, \citenamefont {Hwang}, \citenamefont
  {Huang}, \citenamefont {Dehollain}, \citenamefont {Muhonen}, \citenamefont
  {Simmons}, \citenamefont {Laucht}, \citenamefont {Hudson}, \citenamefont
  {Itoh}, \citenamefont {Morello},\ and\ \citenamefont
  {Dzurak}}]{Veldhorst2015}%
  \BibitemOpen
  \bibfield  {author} {\bibinfo {author} {\bibfnamefont {M.}~\bibnamefont
  {Veldhorst}}, \bibinfo {author} {\bibfnamefont {C.~H.}\ \bibnamefont {Yang}},
  \bibinfo {author} {\bibfnamefont {J.~C.~C.}\ \bibnamefont {Hwang}}, \bibinfo
  {author} {\bibfnamefont {W.}~\bibnamefont {Huang}}, \bibinfo {author}
  {\bibfnamefont {J.~P.}\ \bibnamefont {Dehollain}}, \bibinfo {author}
  {\bibfnamefont {J.~T.}\ \bibnamefont {Muhonen}}, \bibinfo {author}
  {\bibfnamefont {S.}~\bibnamefont {Simmons}}, \bibinfo {author} {\bibfnamefont
  {A.}~\bibnamefont {Laucht}}, \bibinfo {author} {\bibfnamefont {F.~E.}\
  \bibnamefont {Hudson}}, \bibinfo {author} {\bibfnamefont {K.~M.}\
  \bibnamefont {Itoh}}, \bibinfo {author} {\bibfnamefont {A.}~\bibnamefont
  {Morello}}, \ and\ \bibinfo {author} {\bibfnamefont {A.~S.}\ \bibnamefont
  {Dzurak}},\ }\href {http://dx.doi.org/10.1038/nature15263} {\bibfield
  {journal} {\bibinfo  {journal} {Nature}\ }\textbf {\bibinfo {volume} {526}},\
  \bibinfo {pages} {410} (\bibinfo {year} {2015})}\BibitemShut {NoStop}%
\bibitem [{\citenamefont {Zajac}\ \emph {et~al.}(2015)\citenamefont {Zajac},
  \citenamefont {Hazard}, \citenamefont {Mi}, \citenamefont {Wang},\ and\
  \citenamefont {Petta}}]{Zajac2015}%
  \BibitemOpen
  \bibfield  {author} {\bibinfo {author} {\bibfnamefont {D.~M.}\ \bibnamefont
  {Zajac}}, \bibinfo {author} {\bibfnamefont {T.~M.}\ \bibnamefont {Hazard}},
  \bibinfo {author} {\bibfnamefont {X.}~\bibnamefont {Mi}}, \bibinfo {author}
  {\bibfnamefont {K.}~\bibnamefont {Wang}}, \ and\ \bibinfo {author}
  {\bibfnamefont {J.~R.}\ \bibnamefont {Petta}},\ }\href {\doibase
  http://dx.doi.org/10.1063/1.4922249} {\bibfield  {journal} {\bibinfo
  {journal} {Appl. Phys. Lett.}\ }\textbf {\bibinfo {volume} {106}},\ \bibinfo
  {eid} {223507} (\bibinfo {year} {2015})}\BibitemShut {NoStop}%
\bibitem [{\citenamefont {{Otsuka}}\ \emph {et~al.}(2015)\citenamefont
  {{Otsuka}}, \citenamefont {{Nakajima}}, \citenamefont {{Delbecq}},
  \citenamefont {{Amaha}}, \citenamefont {{Yoneda}}, \citenamefont {{Takeda}},
  \citenamefont {{Allison}}, \citenamefont {{Ito}}, \citenamefont {{Sugawara}},
  \citenamefont {{Noiri}}, \citenamefont {{Ludwig}}, \citenamefont {{Wieck}},\
  and\ \citenamefont {{Tarucha}}}]{Otsuka2015}%
  \BibitemOpen
  \bibfield  {author} {\bibinfo {author} {\bibfnamefont {T.}~\bibnamefont
  {{Otsuka}}}, \bibinfo {author} {\bibfnamefont {T.}~\bibnamefont
  {{Nakajima}}}, \bibinfo {author} {\bibfnamefont {M.~R.}\ \bibnamefont
  {{Delbecq}}}, \bibinfo {author} {\bibfnamefont {S.}~\bibnamefont {{Amaha}}},
  \bibinfo {author} {\bibfnamefont {J.}~\bibnamefont {{Yoneda}}}, \bibinfo
  {author} {\bibfnamefont {K.}~\bibnamefont {{Takeda}}}, \bibinfo {author}
  {\bibfnamefont {G.}~\bibnamefont {{Allison}}}, \bibinfo {author}
  {\bibfnamefont {T.}~\bibnamefont {{Ito}}}, \bibinfo {author} {\bibfnamefont
  {R.}~\bibnamefont {{Sugawara}}}, \bibinfo {author} {\bibfnamefont
  {A.}~\bibnamefont {{Noiri}}}, \bibinfo {author} {\bibfnamefont
  {A.}~\bibnamefont {{Ludwig}}}, \bibinfo {author} {\bibfnamefont {A.~D.}\
  \bibnamefont {{Wieck}}}, \ and\ \bibinfo {author} {\bibfnamefont
  {S.}~\bibnamefont {{Tarucha}}},\ }\href@noop {} {\bibfield  {journal}
  {\bibinfo  {journal} {ArXiv e-prints}\ } (\bibinfo {year} {2015})},\ \Eprint
  {http://arxiv.org/abs/1510.02547} {arXiv:1510.02547 [cond-mat.mes-hall]}
  \BibitemShut {NoStop}%
\bibitem [{\citenamefont {DiVincenzo}\ \emph {et~al.}(2000)\citenamefont
  {DiVincenzo}, \citenamefont {Bacon}, \citenamefont {Kempe}, \citenamefont
  {Burkard},\ and\ \citenamefont {Whaley}}]{nature2000}%
  \BibitemOpen
  \bibfield  {author} {\bibinfo {author} {\bibfnamefont {D.~P.}\ \bibnamefont
  {DiVincenzo}}, \bibinfo {author} {\bibfnamefont {D.}~\bibnamefont {Bacon}},
  \bibinfo {author} {\bibfnamefont {J.}~\bibnamefont {Kempe}}, \bibinfo
  {author} {\bibfnamefont {G.}~\bibnamefont {Burkard}}, \ and\ \bibinfo
  {author} {\bibfnamefont {K.~B.}\ \bibnamefont {Whaley}},\ }\href
  {http://dx.doi.org/10.1038/35042541} {\bibfield  {journal} {\bibinfo
  {journal} {Nature}\ }\textbf {\bibinfo {volume} {408}},\ \bibinfo {pages}
  {339} (\bibinfo {year} {2000})}\BibitemShut {NoStop}%
\bibitem [{\citenamefont {Taylor}\ \emph {et~al.}(2013)\citenamefont {Taylor},
  \citenamefont {Srinivasa},\ and\ \citenamefont {Medford}}]{Taylor2013}%
  \BibitemOpen
  \bibfield  {author} {\bibinfo {author} {\bibfnamefont {J.~M.}\ \bibnamefont
  {Taylor}}, \bibinfo {author} {\bibfnamefont {V.}~\bibnamefont {Srinivasa}}, \
  and\ \bibinfo {author} {\bibfnamefont {J.}~\bibnamefont {Medford}},\ }\href
  {\doibase 10.1103/PhysRevLett.111.050502} {\bibfield  {journal} {\bibinfo
  {journal} {Phys. Rev. Lett.}\ }\textbf {\bibinfo {volume} {111}},\ \bibinfo
  {pages} {050502} (\bibinfo {year} {2013})}\BibitemShut {NoStop}%
\bibitem [{\citenamefont {Fei}\ \emph {et~al.}(2015)\citenamefont {Fei},
  \citenamefont {Hung}, \citenamefont {Koh}, \citenamefont {Shim},
  \citenamefont {Coppersmith}, \citenamefont {Hu},\ and\ \citenamefont
  {Friesen}}]{Fei2015}%
  \BibitemOpen
  \bibfield  {author} {\bibinfo {author} {\bibfnamefont {J.}~\bibnamefont
  {Fei}}, \bibinfo {author} {\bibfnamefont {J.-T.}\ \bibnamefont {Hung}},
  \bibinfo {author} {\bibfnamefont {T.~S.}\ \bibnamefont {Koh}}, \bibinfo
  {author} {\bibfnamefont {Y.-P.}\ \bibnamefont {Shim}}, \bibinfo {author}
  {\bibfnamefont {S.~N.}\ \bibnamefont {Coppersmith}}, \bibinfo {author}
  {\bibfnamefont {X.}~\bibnamefont {Hu}}, \ and\ \bibinfo {author}
  {\bibfnamefont {M.}~\bibnamefont {Friesen}},\ }\href {\doibase
  10.1103/PhysRevB.91.205434} {\bibfield  {journal} {\bibinfo  {journal} {Phys.
  Rev. B}\ }\textbf {\bibinfo {volume} {91}},\ \bibinfo {pages} {205434}
  (\bibinfo {year} {2015})}\BibitemShut {NoStop}%
\bibitem [{\citenamefont {Russ}\ and\ \citenamefont
  {Burkard}(2015)}]{Russ2015}%
  \BibitemOpen
  \bibfield  {author} {\bibinfo {author} {\bibfnamefont {M.}~\bibnamefont
  {Russ}}\ and\ \bibinfo {author} {\bibfnamefont {G.}~\bibnamefont {Burkard}},\
  }\href {\doibase 10.1103/PhysRevB.91.235411} {\bibfield  {journal} {\bibinfo
  {journal} {Phys. Rev. B}\ }\textbf {\bibinfo {volume} {91}},\ \bibinfo
  {pages} {235411} (\bibinfo {year} {2015})}\BibitemShut {NoStop}%
\bibitem [{\citenamefont {Doherty}\ and\ \citenamefont
  {Wardrop}(2013)}]{Doherty2013}%
  \BibitemOpen
  \bibfield  {author} {\bibinfo {author} {\bibfnamefont {A.~C.}\ \bibnamefont
  {Doherty}}\ and\ \bibinfo {author} {\bibfnamefont {M.~P.}\ \bibnamefont
  {Wardrop}},\ }\href {\doibase 10.1103/PhysRevLett.111.050503} {\bibfield
  {journal} {\bibinfo  {journal} {Phys. Rev. Lett.}\ }\textbf {\bibinfo
  {volume} {111}},\ \bibinfo {pages} {050503} (\bibinfo {year}
  {2013})}\BibitemShut {NoStop}%
\bibitem [{\citenamefont {Aspelmeyer}\ \emph {et~al.}(2003)\citenamefont
  {Aspelmeyer}, \citenamefont {Jennewein}, \citenamefont {Pfennigbauer},
  \citenamefont {Leeb},\ and\ \citenamefont {Zeilinger}}]{Aspelmeyer2003}%
  \BibitemOpen
  \bibfield  {author} {\bibinfo {author} {\bibfnamefont {M.}~\bibnamefont
  {Aspelmeyer}}, \bibinfo {author} {\bibfnamefont {T.}~\bibnamefont
  {Jennewein}}, \bibinfo {author} {\bibfnamefont {M.}~\bibnamefont
  {Pfennigbauer}}, \bibinfo {author} {\bibfnamefont {W.}~\bibnamefont {Leeb}},
  \ and\ \bibinfo {author} {\bibfnamefont {A.}~\bibnamefont {Zeilinger}},\
  }\href {\doibase 10.1109/JSTQE.2003.820918} {\bibfield  {journal} {\bibinfo
  {journal} {Selected Topics in Quantum Electronics, IEEE Journal of}\ }\textbf
  {\bibinfo {volume} {9}},\ \bibinfo {pages} {1541} (\bibinfo {year}
  {2003})}\BibitemShut {NoStop}%
\bibitem [{\citenamefont {Takeuchi}(2014)}]{Takeuchi2014}%
  \BibitemOpen
  \bibfield  {author} {\bibinfo {author} {\bibfnamefont {S.}~\bibnamefont
  {Takeuchi}},\ }\href {http://stacks.iop.org/1347-4065/53/i=3/a=030101}
  {\bibfield  {journal} {\bibinfo  {journal} {Japanese Journal of Applied
  Physics}\ }\textbf {\bibinfo {volume} {53}},\ \bibinfo {pages} {030101}
  (\bibinfo {year} {2014})}\BibitemShut {NoStop}%
\bibitem [{\citenamefont {Burkard}\ and\ \citenamefont
  {Imamoglu}(2006)}]{Burkard2006}%
  \BibitemOpen
  \bibfield  {author} {\bibinfo {author} {\bibfnamefont {G.}~\bibnamefont
  {Burkard}}\ and\ \bibinfo {author} {\bibfnamefont {A.}~\bibnamefont
  {Imamoglu}},\ }\href {\doibase 10.1103/PhysRevB.74.041307} {\bibfield
  {journal} {\bibinfo  {journal} {Phys. Rev. B}\ }\textbf {\bibinfo {volume}
  {74}},\ \bibinfo {pages} {041307} (\bibinfo {year} {2006})}\BibitemShut
  {NoStop}%
\bibitem [{\citenamefont {Trifunovic}\ \emph {et~al.}(2013)\citenamefont
  {Trifunovic}, \citenamefont {Pedrocchi},\ and\ \citenamefont
  {Loss}}]{Trifunovic2013}%
  \BibitemOpen
  \bibfield  {author} {\bibinfo {author} {\bibfnamefont {L.}~\bibnamefont
  {Trifunovic}}, \bibinfo {author} {\bibfnamefont {F.~L.}\ \bibnamefont
  {Pedrocchi}}, \ and\ \bibinfo {author} {\bibfnamefont {D.}~\bibnamefont
  {Loss}},\ }\href {\doibase 10.1103/PhysRevX.3.041023} {\bibfield  {journal}
  {\bibinfo  {journal} {Phys. Rev. X}\ }\textbf {\bibinfo {volume} {3}},\
  \bibinfo {pages} {041023} (\bibinfo {year} {2013})}\BibitemShut {NoStop}%
\bibitem [{\citenamefont {Dobrovitski}\ \emph {et~al.}(2013)\citenamefont
  {Dobrovitski}, \citenamefont {Fuchs}, \citenamefont {Falk}, \citenamefont
  {Santori},\ and\ \citenamefont {Awschalom}}]{Dobrovitski2013}%
  \BibitemOpen
  \bibfield  {author} {\bibinfo {author} {\bibfnamefont {V.}~\bibnamefont
  {Dobrovitski}}, \bibinfo {author} {\bibfnamefont {G.}~\bibnamefont {Fuchs}},
  \bibinfo {author} {\bibfnamefont {A.}~\bibnamefont {Falk}}, \bibinfo {author}
  {\bibfnamefont {C.}~\bibnamefont {Santori}}, \ and\ \bibinfo {author}
  {\bibfnamefont {D.}~\bibnamefont {Awschalom}},\ }\href {\doibase
  10.1146/annurev-conmatphys-030212-184238} {\bibfield  {journal} {\bibinfo
  {journal} {Annual Review of Condensed Matter Physics}\ }\textbf {\bibinfo
  {volume} {4}},\ \bibinfo {pages} {23} (\bibinfo {year} {2013})}\BibitemShut
  {NoStop}%
\bibitem [{\citenamefont {{Burkard}}\ and\ \citenamefont
  {{Awschalom}}(2014)}]{Burkard2014}%
  \BibitemOpen
  \bibfield  {author} {\bibinfo {author} {\bibfnamefont {G.}~\bibnamefont
  {{Burkard}}}\ and\ \bibinfo {author} {\bibfnamefont {D.~D.}\ \bibnamefont
  {{Awschalom}}},\ }\href@noop {} {\bibfield  {journal} {\bibinfo  {journal}
  {ArXiv e-prints}\ } (\bibinfo {year} {2014})},\ \Eprint
  {http://arxiv.org/abs/1402.6351} {arXiv:1402.6351 [cond-mat.mes-hall]}
  \BibitemShut {NoStop}%
\bibitem [{\citenamefont {Gao}\ \emph {et~al.}(2015)\citenamefont {Gao},
  \citenamefont {Imamoglu}, \citenamefont {Bernien},\ and\ \citenamefont
  {Hanson}}]{Gao2015}%
  \BibitemOpen
  \bibfield  {author} {\bibinfo {author} {\bibfnamefont {W.~B.}\ \bibnamefont
  {Gao}}, \bibinfo {author} {\bibfnamefont {A.}~\bibnamefont {Imamoglu}},
  \bibinfo {author} {\bibfnamefont {H.}~\bibnamefont {Bernien}}, \ and\
  \bibinfo {author} {\bibfnamefont {R.}~\bibnamefont {Hanson}},\ }\href
  {http://dx.doi.org/10.1038/nphoton.2015.58} {\bibfield  {journal} {\bibinfo
  {journal} {Nat Photon}\ }\textbf {\bibinfo {volume} {9}},\ \bibinfo {pages}
  {363} (\bibinfo {year} {2015})}\BibitemShut {NoStop}%
\bibitem [{\citenamefont {Xiang}\ \emph {et~al.}(2013)\citenamefont {Xiang},
  \citenamefont {Ashhab}, \citenamefont {You},\ and\ \citenamefont
  {Nori}}]{Xiang2014}%
  \BibitemOpen
  \bibfield  {author} {\bibinfo {author} {\bibfnamefont {Z.-L.}\ \bibnamefont
  {Xiang}}, \bibinfo {author} {\bibfnamefont {S.}~\bibnamefont {Ashhab}},
  \bibinfo {author} {\bibfnamefont {J.~Q.}\ \bibnamefont {You}}, \ and\
  \bibinfo {author} {\bibfnamefont {F.}~\bibnamefont {Nori}},\ }\href {\doibase
  10.1103/RevModPhys.85.623} {\bibfield  {journal} {\bibinfo  {journal} {Rev.
  Mod. Phys.}\ }\textbf {\bibinfo {volume} {85}},\ \bibinfo {pages} {623–653}
  (\bibinfo {year} {2013})}\BibitemShut {NoStop}%
\bibitem [{\citenamefont {Hu}\ \emph {et~al.}(2012)\citenamefont {Hu},
  \citenamefont {Liu},\ and\ \citenamefont {Nori}}]{Hu2012}%
  \BibitemOpen
  \bibfield  {author} {\bibinfo {author} {\bibfnamefont {X.}~\bibnamefont
  {Hu}}, \bibinfo {author} {\bibfnamefont {Y.-x.}\ \bibnamefont {Liu}}, \ and\
  \bibinfo {author} {\bibfnamefont {F.}~\bibnamefont {Nori}},\ }\href {\doibase
  10.1103/PhysRevB.86.035314} {\bibfield  {journal} {\bibinfo  {journal} {Phys.
  Rev. B}\ }\textbf {\bibinfo {volume} {86}},\ \bibinfo {pages} {035314}
  (\bibinfo {year} {2012})}\BibitemShut {NoStop}%
\bibitem [{\citenamefont {Jin}\ \emph {et~al.}(2012)\citenamefont {Jin},
  \citenamefont {Marthaler}, \citenamefont {Shnirman},\ and\ \citenamefont
  {Sch\"on}}]{Jin2012}%
  \BibitemOpen
  \bibfield  {author} {\bibinfo {author} {\bibfnamefont {P.-Q.}\ \bibnamefont
  {Jin}}, \bibinfo {author} {\bibfnamefont {M.}~\bibnamefont {Marthaler}},
  \bibinfo {author} {\bibfnamefont {A.}~\bibnamefont {Shnirman}}, \ and\
  \bibinfo {author} {\bibfnamefont {G.}~\bibnamefont {Sch\"on}},\ }\href
  {\doibase 10.1103/PhysRevLett.108.190506} {\bibfield  {journal} {\bibinfo
  {journal} {Phys. Rev. Lett.}\ }\textbf {\bibinfo {volume} {108}},\ \bibinfo
  {pages} {190506} (\bibinfo {year} {2012})}\BibitemShut {NoStop}%
\bibitem [{\citenamefont {Petersson}\ \emph {et~al.}(2012)\citenamefont
  {Petersson}, \citenamefont {McFaul}, \citenamefont {Schroer}, \citenamefont
  {Jung}, \citenamefont {Taylor}, \citenamefont {Houck},\ and\ \citenamefont
  {Petta}}]{Petersson2012}%
  \BibitemOpen
  \bibfield  {author} {\bibinfo {author} {\bibfnamefont {K.~D.}\ \bibnamefont
  {Petersson}}, \bibinfo {author} {\bibfnamefont {L.~W.}\ \bibnamefont
  {McFaul}}, \bibinfo {author} {\bibfnamefont {M.~D.}\ \bibnamefont {Schroer}},
  \bibinfo {author} {\bibfnamefont {M.}~\bibnamefont {Jung}}, \bibinfo {author}
  {\bibfnamefont {J.~M.}\ \bibnamefont {Taylor}}, \bibinfo {author}
  {\bibfnamefont {A.~A.}\ \bibnamefont {Houck}}, \ and\ \bibinfo {author}
  {\bibfnamefont {J.~R.}\ \bibnamefont {Petta}},\ }\href {\doibase
  10.1038/nature11559} {\bibfield  {journal} {\bibinfo  {journal} {Nature}\
  }\textbf {\bibinfo {volume} {490}},\ \bibinfo {pages} {380} (\bibinfo {year}
  {2012})}\BibitemShut {NoStop}%
\bibitem [{\citenamefont {Toida}\ \emph {et~al.}(2013)\citenamefont {Toida},
  \citenamefont {Nakajima},\ and\ \citenamefont {Komiyama}}]{Toida2013}%
  \BibitemOpen
  \bibfield  {author} {\bibinfo {author} {\bibfnamefont {H.}~\bibnamefont
  {Toida}}, \bibinfo {author} {\bibfnamefont {T.}~\bibnamefont {Nakajima}}, \
  and\ \bibinfo {author} {\bibfnamefont {S.}~\bibnamefont {Komiyama}},\ }\href
  {\doibase 10.1103/PhysRevLett.110.066802} {\bibfield  {journal} {\bibinfo
  {journal} {Phys. Rev. Lett.}\ }\textbf {\bibinfo {volume} {110}},\ \bibinfo
  {pages} {066802} (\bibinfo {year} {2013})}\BibitemShut {NoStop}%
\bibitem [{\citenamefont {Wallraff}\ \emph {et~al.}(2013)\citenamefont
  {Wallraff}, \citenamefont {Stockklauser}, \citenamefont {Ihn}, \citenamefont
  {Petta},\ and\ \citenamefont {Blais}}]{Wallraff2013}%
  \BibitemOpen
  \bibfield  {author} {\bibinfo {author} {\bibfnamefont {A.}~\bibnamefont
  {Wallraff}}, \bibinfo {author} {\bibfnamefont {A.}~\bibnamefont
  {Stockklauser}}, \bibinfo {author} {\bibfnamefont {T.}~\bibnamefont {Ihn}},
  \bibinfo {author} {\bibfnamefont {J.~R.}\ \bibnamefont {Petta}}, \ and\
  \bibinfo {author} {\bibfnamefont {A.}~\bibnamefont {Blais}},\ }\href
  {\doibase 10.1103/PhysRevLett.111.249701} {\bibfield  {journal} {\bibinfo
  {journal} {Phys. Rev. Lett.}\ }\textbf {\bibinfo {volume} {111}},\ \bibinfo
  {pages} {249701} (\bibinfo {year} {2013})}\BibitemShut {NoStop}%
\bibitem [{\citenamefont {Liu}\ \emph {et~al.}(2014)\citenamefont {Liu},
  \citenamefont {Petersson}, \citenamefont {Stehlik}, \citenamefont {Taylor},\
  and\ \citenamefont {Petta}}]{Liu2014}%
  \BibitemOpen
  \bibfield  {author} {\bibinfo {author} {\bibfnamefont {Y.-Y.}\ \bibnamefont
  {Liu}}, \bibinfo {author} {\bibfnamefont {K.~D.}\ \bibnamefont {Petersson}},
  \bibinfo {author} {\bibfnamefont {J.}~\bibnamefont {Stehlik}}, \bibinfo
  {author} {\bibfnamefont {J.~M.}\ \bibnamefont {Taylor}}, \ and\ \bibinfo
  {author} {\bibfnamefont {J.~R.}\ \bibnamefont {Petta}},\ }\href {\doibase
  10.1103/PhysRevLett.113.036801} {\bibfield  {journal} {\bibinfo  {journal}
  {Phys. Rev. Lett.}\ }\textbf {\bibinfo {volume} {113}},\ \bibinfo {pages}
  {036801} (\bibinfo {year} {2014})}\BibitemShut {NoStop}%
\bibitem [{\citenamefont {Frey}\ \emph {et~al.}(2012)\citenamefont {Frey},
  \citenamefont {Leek}, \citenamefont {Beck}, \citenamefont {Blais},
  \citenamefont {Ihn}, \citenamefont {Ensslin},\ and\ \citenamefont
  {Wallraff}}]{Frey2012}%
  \BibitemOpen
  \bibfield  {author} {\bibinfo {author} {\bibfnamefont {T.}~\bibnamefont
  {Frey}}, \bibinfo {author} {\bibfnamefont {P.~J.}\ \bibnamefont {Leek}},
  \bibinfo {author} {\bibfnamefont {M.}~\bibnamefont {Beck}}, \bibinfo {author}
  {\bibfnamefont {A.}~\bibnamefont {Blais}}, \bibinfo {author} {\bibfnamefont
  {T.}~\bibnamefont {Ihn}}, \bibinfo {author} {\bibfnamefont {K.}~\bibnamefont
  {Ensslin}}, \ and\ \bibinfo {author} {\bibfnamefont {A.}~\bibnamefont
  {Wallraff}},\ }\href {\doibase 10.1103/PhysRevLett.108.046807} {\bibfield
  {journal} {\bibinfo  {journal} {Phys. Rev. Lett.}\ }\textbf {\bibinfo
  {volume} {108}},\ \bibinfo {pages} {046807} (\bibinfo {year}
  {2012})}\BibitemShut {NoStop}%
\bibitem [{\citenamefont {Basset}\ \emph {et~al.}(2013)\citenamefont {Basset},
  \citenamefont {Jarausch}, \citenamefont {Stockklauser}, \citenamefont {Frey},
  \citenamefont {Reichl}, \citenamefont {Wegscheider}, \citenamefont {Ihn},
  \citenamefont {Ensslin},\ and\ \citenamefont {Wallraff}}]{Basset2013}%
  \BibitemOpen
  \bibfield  {author} {\bibinfo {author} {\bibfnamefont {J.}~\bibnamefont
  {Basset}}, \bibinfo {author} {\bibfnamefont {D.-D.}\ \bibnamefont
  {Jarausch}}, \bibinfo {author} {\bibfnamefont {A.}~\bibnamefont
  {Stockklauser}}, \bibinfo {author} {\bibfnamefont {T.}~\bibnamefont {Frey}},
  \bibinfo {author} {\bibfnamefont {C.}~\bibnamefont {Reichl}}, \bibinfo
  {author} {\bibfnamefont {W.}~\bibnamefont {Wegscheider}}, \bibinfo {author}
  {\bibfnamefont {T.~M.}\ \bibnamefont {Ihn}}, \bibinfo {author} {\bibfnamefont
  {K.}~\bibnamefont {Ensslin}}, \ and\ \bibinfo {author} {\bibfnamefont
  {A.}~\bibnamefont {Wallraff}},\ }\href {\doibase 10.1103/PhysRevB.88.125312}
  {\bibfield  {journal} {\bibinfo  {journal} {Phys. Rev. B}\ }\textbf {\bibinfo
  {volume} {88}},\ \bibinfo {pages} {125312} (\bibinfo {year}
  {2013})}\BibitemShut {NoStop}%
\bibitem [{\citenamefont {Cohen-Tannoudji}\ \emph {et~al.}(2004)\citenamefont
  {Cohen-Tannoudji}, \citenamefont {Dupont-Roc},\ and\ \citenamefont
  {Grynberg}}]{cohen1997}%
  \BibitemOpen
  \bibfield  {author} {\bibinfo {author} {\bibfnamefont {C.}~\bibnamefont
  {Cohen-Tannoudji}}, \bibinfo {author} {\bibfnamefont {J.}~\bibnamefont
  {Dupont-Roc}}, \ and\ \bibinfo {author} {\bibfnamefont {G.}~\bibnamefont
  {Grynberg}},\ }\href {\doibase 10.1002/9783527618422} {\emph {\bibinfo
  {title} {Photons and Atoms: Introduction to Quantum Electrodynamics}}}\
  (\bibinfo  {publisher} {Wiley-VCH},\ \bibinfo {year} {2004})\BibitemShut
  {NoStop}%
\bibitem [{\citenamefont {Hung}\ \emph {et~al.}(2014)\citenamefont {Hung},
  \citenamefont {Fei}, \citenamefont {Friesen},\ and\ \citenamefont
  {Hu}}]{Hung2014}%
  \BibitemOpen
  \bibfield  {author} {\bibinfo {author} {\bibfnamefont {J.-T.}\ \bibnamefont
  {Hung}}, \bibinfo {author} {\bibfnamefont {J.}~\bibnamefont {Fei}}, \bibinfo
  {author} {\bibfnamefont {M.}~\bibnamefont {Friesen}}, \ and\ \bibinfo
  {author} {\bibfnamefont {X.}~\bibnamefont {Hu}},\ }\href {\doibase
  10.1103/PhysRevB.90.045308} {\bibfield  {journal} {\bibinfo  {journal} {Phys.
  Rev. B}\ }\textbf {\bibinfo {volume} {90}},\ \bibinfo {pages} {045308}
  (\bibinfo {year} {2014})}\BibitemShut {NoStop}%
\bibitem [{\citenamefont {Marzari}\ \emph {et~al.}(2012)\citenamefont
  {Marzari}, \citenamefont {Mostofi}, \citenamefont {Yates}, \citenamefont
  {Souza},\ and\ \citenamefont {Vanderbilt}}]{Marzari2012}%
  \BibitemOpen
  \bibfield  {author} {\bibinfo {author} {\bibfnamefont {N.}~\bibnamefont
  {Marzari}}, \bibinfo {author} {\bibfnamefont {A.~A.}\ \bibnamefont
  {Mostofi}}, \bibinfo {author} {\bibfnamefont {J.~R.}\ \bibnamefont {Yates}},
  \bibinfo {author} {\bibfnamefont {I.}~\bibnamefont {Souza}}, \ and\ \bibinfo
  {author} {\bibfnamefont {D.}~\bibnamefont {Vanderbilt}},\ }\href {\doibase
  10.1103/RevModPhys.84.1419} {\bibfield  {journal} {\bibinfo  {journal} {Rev.
  Mod. Phys.}\ }\textbf {\bibinfo {volume} {84}},\ \bibinfo {pages} {1419}
  (\bibinfo {year} {2012})}\BibitemShut {NoStop}%
\bibitem [{\citenamefont {Milivojevi\'{c}}\ and\ \citenamefont
  {Stepanenko}()}]{Milkojevic2015}%
  \BibitemOpen
  \bibfield  {author} {\bibinfo {author} {\bibfnamefont {M.}~\bibnamefont
  {Milivojevi\'{c}}}\ and\ \bibinfo {author} {\bibfnamefont {D.}~\bibnamefont
  {Stepanenko}},\ }\href@noop {} {\bibinfo  {journal} {unpublished}\
  }\BibitemShut {NoStop}%
\bibitem [{\citenamefont {Burkard}\ \emph {et~al.}(1999)\citenamefont
  {Burkard}, \citenamefont {Loss},\ and\ \citenamefont
  {DiVincenzo}}]{Burkard1999}%
  \BibitemOpen
\bibfield  {journal} {  }\bibfield  {author} {\bibinfo {author} {\bibfnamefont
  {G.}~\bibnamefont {Burkard}}, \bibinfo {author} {\bibfnamefont
  {D.}~\bibnamefont {Loss}}, \ and\ \bibinfo {author} {\bibfnamefont {D.~P.}\
  \bibnamefont {DiVincenzo}},\ }\href {\doibase 10.1103/PhysRevB.59.2070}
  {\bibfield  {journal} {\bibinfo  {journal} {Phys. Rev. B}\ }\textbf {\bibinfo
  {volume} {59}},\ \bibinfo {pages} {2070} (\bibinfo {year}
  {1999})}\BibitemShut {NoStop}%
\bibitem [{\citenamefont {Imamoglu}\ \emph {et~al.}(1999)\citenamefont
  {Imamoglu}, \citenamefont {Awschalom}, \citenamefont {Burkard}, \citenamefont
  {DiVincenzo}, \citenamefont {Loss}, \citenamefont {Sherwin},\ and\
  \citenamefont {Small}}]{Imamoglu1999}%
  \BibitemOpen
  \bibfield  {author} {\bibinfo {author} {\bibfnamefont {A.}~\bibnamefont
  {Imamoglu}}, \bibinfo {author} {\bibfnamefont {D.~D.}\ \bibnamefont
  {Awschalom}}, \bibinfo {author} {\bibfnamefont {G.}~\bibnamefont {Burkard}},
  \bibinfo {author} {\bibfnamefont {D.~P.}\ \bibnamefont {DiVincenzo}},
  \bibinfo {author} {\bibfnamefont {D.}~\bibnamefont {Loss}}, \bibinfo {author}
  {\bibfnamefont {M.}~\bibnamefont {Sherwin}}, \ and\ \bibinfo {author}
  {\bibfnamefont {A.}~\bibnamefont {Small}},\ }\href {\doibase
  10.1103/PhysRevLett.83.4204} {\bibfield  {journal} {\bibinfo  {journal}
  {Phys. Rev. Lett.}\ }\textbf {\bibinfo {volume} {83}},\ \bibinfo {pages}
  {4204} (\bibinfo {year} {1999})}\BibitemShut {NoStop}%
\bibitem [{\citenamefont {Tanamoto}\ \emph {et~al.}(2008)\citenamefont
  {Tanamoto}, \citenamefont {Maruyama}, \citenamefont {Liu}, \citenamefont
  {Hu},\ and\ \citenamefont {Nori}}]{Tanamota2008}%
  \BibitemOpen
  \bibfield  {author} {\bibinfo {author} {\bibfnamefont {T.}~\bibnamefont
  {Tanamoto}}, \bibinfo {author} {\bibfnamefont {K.}~\bibnamefont {Maruyama}},
  \bibinfo {author} {\bibfnamefont {Y.-x.}\ \bibnamefont {Liu}}, \bibinfo
  {author} {\bibfnamefont {X.}~\bibnamefont {Hu}}, \ and\ \bibinfo {author}
  {\bibfnamefont {F.}~\bibnamefont {Nori}},\ }\href {\doibase
  10.1103/PhysRevA.78.062313} {\bibfield  {journal} {\bibinfo  {journal} {Phys.
  Rev. A}\ }\textbf {\bibinfo {volume} {78}},\ \bibinfo {pages} {062313}
  (\bibinfo {year} {2008})}\BibitemShut {NoStop}%
\bibitem [{\citenamefont {Culcer}\ \emph {et~al.}(2009)\citenamefont {Culcer},
  \citenamefont {Cywi\ifmmode~\acute{n}\else \'{n}\fi{}ski}, \citenamefont
  {Li}, \citenamefont {Hu},\ and\ \citenamefont {Das~Sarma}}]{Culcer2009}%
  \BibitemOpen
  \bibfield  {author} {\bibinfo {author} {\bibfnamefont {D.}~\bibnamefont
  {Culcer}}, \bibinfo {author} {\bibfnamefont {L.}~\bibnamefont
  {Cywi\ifmmode~\acute{n}\else \'{n}\fi{}ski}}, \bibinfo {author}
  {\bibfnamefont {Q.}~\bibnamefont {Li}}, \bibinfo {author} {\bibfnamefont
  {X.}~\bibnamefont {Hu}}, \ and\ \bibinfo {author} {\bibfnamefont
  {S.}~\bibnamefont {Das~Sarma}},\ }\href {\doibase 10.1103/PhysRevB.80.205302}
  {\bibfield  {journal} {\bibinfo  {journal} {Phys. Rev. B}\ }\textbf {\bibinfo
  {volume} {80}},\ \bibinfo {pages} {205302} (\bibinfo {year}
  {2009})}\BibitemShut {NoStop}%
\bibitem [{\citenamefont {Culcer}\ \emph {et~al.}(2010)\citenamefont {Culcer},
  \citenamefont {Cywi\ifmmode~\acute{n}\else \'{n}\fi{}ski}, \citenamefont
  {Li}, \citenamefont {Hu},\ and\ \citenamefont {Das~Sarma}}]{Culcer2010}%
  \BibitemOpen
  \bibfield  {author} {\bibinfo {author} {\bibfnamefont {D.}~\bibnamefont
  {Culcer}}, \bibinfo {author} {\bibfnamefont {L.}~\bibnamefont
  {Cywi\ifmmode~\acute{n}\else \'{n}\fi{}ski}}, \bibinfo {author}
  {\bibfnamefont {Q.}~\bibnamefont {Li}}, \bibinfo {author} {\bibfnamefont
  {X.}~\bibnamefont {Hu}}, \ and\ \bibinfo {author} {\bibfnamefont
  {S.}~\bibnamefont {Das~Sarma}},\ }\href {\doibase 10.1103/PhysRevB.82.155312}
  {\bibfield  {journal} {\bibinfo  {journal} {Phys. Rev. B}\ }\textbf {\bibinfo
  {volume} {82}},\ \bibinfo {pages} {155312} (\bibinfo {year}
  {2010})}\BibitemShut {NoStop}%
\bibitem [{\citenamefont {Rohling}\ and\ \citenamefont
  {Burkard}(2012)}]{Rohling2012}%
  \BibitemOpen
  \bibfield  {author} {\bibinfo {author} {\bibfnamefont {N.}~\bibnamefont
  {Rohling}}\ and\ \bibinfo {author} {\bibfnamefont {G.}~\bibnamefont
  {Burkard}},\ }\href {http://stacks.iop.org/1367-2630/14/i=8/a=083008}
  {\bibfield  {journal} {\bibinfo  {journal} {New Journal of Physics}\ }\textbf
  {\bibinfo {volume} {14}},\ \bibinfo {pages} {083008} (\bibinfo {year}
  {2012})}\BibitemShut {NoStop}%
\bibitem [{\citenamefont {Rohling}\ \emph {et~al.}(2014)\citenamefont
  {Rohling}, \citenamefont {Russ},\ and\ \citenamefont
  {Burkard}}]{Rohling2014}%
  \BibitemOpen
  \bibfield  {author} {\bibinfo {author} {\bibfnamefont {N.}~\bibnamefont
  {Rohling}}, \bibinfo {author} {\bibfnamefont {M.}~\bibnamefont {Russ}}, \
  and\ \bibinfo {author} {\bibfnamefont {G.}~\bibnamefont {Burkard}},\ }\href
  {\doibase 10.1103/PhysRevLett.113.176801} {\bibfield  {journal} {\bibinfo
  {journal} {Phys. Rev. Lett.}\ }\textbf {\bibinfo {volume} {113}},\ \bibinfo
  {pages} {176801} (\bibinfo {year} {2014})}\BibitemShut {NoStop}%
\bibitem [{\citenamefont {Datta}(1999)}]{datta1999}%
  \BibitemOpen
  \bibfield  {author} {\bibinfo {author} {\bibfnamefont {S.}~\bibnamefont
  {Datta}},\ }\href@noop {} {\emph {\bibinfo {title} {Electronic Transport in
  Mesoscopic Systems}}},\ Cambridge Studies in Semiconductor Physics and
  Microelectronic Engineering\ (\bibinfo  {publisher} {Cambridge University
  Press},\ \bibinfo {year} {1999})\BibitemShut {NoStop}%
\end{thebibliography}%

\end{document}